\documentclass[a4paper,12pt]{article}

\usepackage[german,american]{babel}

\usepackage{times}
\usepackage{latexsym}
\usepackage{ulem}

\usepackage{bm}
\usepackage{amsfonts}
\usepackage{amssymb}
\usepackage{amsmath}
\usepackage{mathtools}

\usepackage[OMLmathsfit]{isomath}

\usepackage[table,xcdraw]{xcolor}
\definecolor{darkgreen}{rgb}{0,0.6,0}
\definecolor{gray}{rgb}{0.5,0.5,0.5}
\definecolor{mauve}{rgb}{0.58,0,0.82}

\usepackage{graphicx}

\graphicspath{{Figures/}}

\usepackage{float}

\usepackage[font=small,labelfont=rm]{caption}
\usepackage{subcaption}

\DeclareCaptionLabelFormat{mylabel}{#1 #2.\hspace{1mm}}
\usepackage{verbatim}
\usepackage[colorlinks=true, allcolors=blue]{hyperref}
\usepackage{cleveref}

\usepackage[sort&compress]{natbib}

\bibpunct[]{(}{)}{;}{a}{,}{,}
\usepackage{lineno}

\begin{document}
\thispagestyle{empty}

\begin{center}
%\color{red}
\textbf{\Large A physics-encoded Fourier neural operator approach 
for surrogate modeling of divergence-free stress fields in solids} 
\color{black}
\\[5mm]
\textrm{Mohammad S. Khorrami$^{1,*}$, 
Pawan Goyal$^{2}$,
Jaber R. Mianroodi$^{1}$,\\
Bob Svendsen$^{1,3}$, Peter Benner$^{2}$, Dierk Raabe$^{1}$}
\\[3mm]
\textrm{\footnotesize 
${}^{1}$Microstructure Physics and Alloy Design,\\
Max-Planck Institute for Sustainable Materials, 
D\"usseldorf, Germany
\\[2mm]
${}^{2}$Computational Methods in Systems and Control Theory,\\ 
Max-Planck Institute for Dynamics of Complex Technical Systems, 
Magdeburg, Germany
\\[2mm]
${}^{3}$Material Mechanics, RWTH Aachen University, Aachen, Germany
\\[1mm]
${}^{*}$Corresponding Author: m.khorrami@mpie.de} 
\end{center}

\begin{abstract}

The purpose of the current work is the development of a so-called 
physics-encoded Fourier neural operator (PeFNO) for surrogate 
modeling of the quasi-static equilibrium stress field in solids. 
Rather than accounting for constraints from physics in the loss 
function as done in the (now standard) physics-informed approach,  
the physics-encoded approach incorporates or "encodes" such 
constraints directly into the network or operator architecture. 
As a result, in contrast to the physics-informed approach in which 
only training is physically constrained, both training and output are 
physically constrained in the physics-encoded approach. For the 
current constraint of divergence-free stress, a novel encoding approach 
based on a stress potential is proposed. 

As a "proof-of-concept" example application of the proposed PeFNO, 
a heterogeneous polycrystalline material consisting of isotropic elastic grains 
subject to uniaxial extension is considered. Stress field data for 
training are obtained from the numerical solution of a corresponding 
boundary-value problem for quasi-static mechanical equilibrium. 
This data is also employed to train an analogous physics-guided FNO 
(PgFNO) and physics-informed FNO (PiFNO) for comparison. As confirmed by 
this comparison and as expected on the basis of their differences, 
the output of the trained PeFNO is significantly 
more accurate in satisfying mechanical equilibrium than the output 
of either the trained PgFNO or the trained PiFNO. 

\end{abstract}


\section{Introduction}
\label{sec:intro}

Given the inherent lack and sparsity of data for most physical phenomena 
as well as their complexity, the development of surrogate numerical models 
for these phenomena based on (artificial) neural networks (NNs) 
often includes constraints from physics to improve model robustness. 
As discussed recently for example by \cite{Faroughi2024}, 
the resulting physics-constrained NNs are currently of three 
types:~(i) physics-guided (PgNNs), 
(ii) physics-informed (PiNNs), 
and (iii) physics-encoded (PeNNs). 
For all three types, the training and testing data are constrained to be physical; 
PiNNs and PeNNs are based in addition on further physical constraints. 
In the case of PiNNs, these constraints are incorporated into the 
loss function (for training/testing), whereas for PeNNs, they are 
incorporated into ("encoded" in) the network architecture. 
The basic rational or motivation behind the idea of constrained 
training and / or output is to reduce the sensitivity of trained 
NN/NO accuracy to the size of the data set. 

By far the most common of these three types are PgNNs and PiNNs. 
PiNNs in particular have been developed for various problems in 
science and engineering \citep{cuomo2022scientific}, such as 
computational fluid dynamics \citep{cai2021physics, mahmoudabadbozchelou2022nn} 
or heat transfer \citep{cai2021physics-Heat, xu2023physics, oommen2022solving}. 
Additional examples of PgNNs and PiNNs for computational fluid flow 
are discussed by \citet[][, Tables 2, 4, 5]{Faroughi2024}. 
Analogously, a number of PgNNs and PiNNs have been developed for surrogate 
numerical modeling in computational solid mechanics 
\citep{goswami2020transfer, abueidda2021meshless, 
mianroodi2021teaching, kumar2022machine, 
wessels2022computational, bai2023physics, diao2023solving, 
khorrami2023artificial, yang2023machine}; see also 
\citet[][, Tables 3, 6, 7]{Faroughi2024}. 

Among the types of physics-constrained NNs mentioned above, 
PeNNs are the most challenging to develop \cite[e.g.,][, \S 4]{Faroughi2024} 
and consequently the least common in the literature. Advantages 
of PeNNs include much less sensitivity to data sparsity than exhibited 
by PgNNs and PiNNs. 
Prominent examples of PeNNs include 
(i) physics-encoded recurrent convolutional neural network (PeRCNN) 
and (ii) neural ordinary differential equations (NeuralODE). 
NeuralPDE \citep{Dulny2021} represents an extension of NeuralODE 
to PDEs by combining the method of lines and NeuralODE in a 
multilayer convolutional NN. 
Being dependent on the method of lines, NeuralPDE is not directly 
applicable to certain kinds of PDEs, e.g., elliptical second-order PDEs. 

%\color{red}

Methodolgically more closely related to the current work 
is that of \cite{RichterPowell2022}, who developed a PeNN for 
divergence-free vector fields 
(e.g., the velocity field in the case of incompressible flow) 
in computational fluid dynamics. 
In contrast to the current approach suggested here, 
their very interesting approach 
(i) is limited to vector fields, 
(ii) is based on the Hodge decomposition of differential 1-forms, 
and (iii) employs function approximation (i.e., NNs) and 
automatic differentiation. 

\color{black}

In computational solid mechanics, PeNNs are rare; for example, 
\citet[][, Table 8]{Faroughi2024} document no such PeNN. 
In fact, to our knowledge, the only PeNN for computational solid 
mechanics in the literature is the so-called constitutive artificial NN (CANN) 
of \cite{Linka2023} for constitutive modeling of isotropic elastic solids. 
More specifically, CANN represents a PeNN for finite-deformation 
isotropic hyperelasticity which exploits tensor invariants and so-called 
structure tensors \cite[e.g.,][]{Svendsen1994,Zheng1994Review} to 
parameterize the network, and encodes physically relevant mathematical 
properties such as polyconvexity in the network architecture. 

Rather than the function-based approximation of neural networks, 
so-called neural operators \cite[NOs:~e.g.,][, \S 5]{Faroughi2024} 
employ approximation based on functionals. 
Advantages of NOs over NNs include for example insensitivity to 
numerical resolution. 
Examples of NOs include deep operator networks (DeepONets) \citep{Lu2019,Lu2021} 
and Fourier neural operators (FNOs) \citep{Li2021,Kovachki2023}. 
In particular, PgNOs and PiNOs have been developed for surrogate numerical 
modeling in computational solid mechanics, including 
heterogeneous materials \citep{kapoor2022comparison}, 
linear elastic fracture mechanics 
\citep{goswami2022fracture, goswami2023physics}, or even 
production by three-dimensional printing \citep{rashid2022learning}. 

The purpose of the current work is the development of a PeFNO for 
surrogate numerical modeling of quasi-static mechanical equilibrium 
and the corresponding divergence-free stress field. This is based on 
generalization of the network/operator architecture and especially 
its output layer to encode divergence-free stress with the help of a 
stress potential. In particular, the architecture of the current PeFNO 
represents a corresponding direct generalization of the FNO from \cite{Li2021}. 
%\color{red} 
To our knowledge, neither a PeNN, nor a PeNO, of this type for 
surrogate numerical modeling of the quasi-static equilibrium stress 
field has appeared in the literature up to this point. 
\color{black}

The paper is organized as follows. 
Section \ref{sec:PcNA} begins with a treatment and discussion of aspects 
common to the development of physics-constrained NNs 
and NOs for the quasi-static equilibrium stress field in solids. 
%\color{red}
To facilitate this and comparison with the literature, the term 
"neural approximation" (NA) is employed for both NNs and NOs 
in this work \cite[this term was also used for example by][]{RichterPowell2022}. 
\color{black} 
Common aspects of NAs include in particular the (i) architecture 
and (ii) loss function. 
In this context, the current approach to encode quasi-static mechanical 
equilibrium (i.e., divergence-free stress) in the NA architecture via a 
stress potential is treated in detail. To demonstrate the capabilities 
of the resulting PeFNO for divergence-free stress, it is applied in 
Section \ref{sec:ExaCom} to surrogate modeling of the stress field in a polycrystalline 
solid and compared with analogous Pg- and PiFNOs. 
Synthetic stress field data for training is generated via numerical solution of 
a physical boundary value problem (BVP) for quasi-static mechanical 
equilibrium in polycrystals consisting of isotropic elastic grains subject to 
uniaxial extension. 
The paper ends in Section \ref{sec:OutSum} with a summary and outlook. 
%\color{red}
To address a broader audience, the treatment and presentation to follow 
is mostly conceptual and descriptive in character. Because of their 
scientific importance, as well as for the more technically incline reader, 
the mathematical and computational details relevant to the current 
PeFNO for surrogate modeling of divergence-free stress are summarized 
in the Supplementary Information. 
\color{black}

\section{Physics-constrained NAs for the stress field}
\label{sec:PcNA}

For brevity, the treatment to follow is restricted to the case of 
geometric-non-linear solid mechanics \cite[e.g.,][]{Tru65,Silhavy1997}. 
%\color{red}
The corresponding treatment for the geometric-linear case 
is briefly summarized in the Supplementary Information. 
\color{black} 
In the non-linear case, the relevant stress measure is the first 
Piola-Kirchhoff stress, represented here for simplicity by the matrix 
of its Cartesian components 
\begin{equation}
\mathbf{P}
=\left\lbrack
\begin{array}{ccc}
P_{11}&P_{12}&P_{13}
\\
P_{21}&P_{22}&P_{23}
\\
P_{31}&P_{32}&P_{33}
\end{array}
\right\rbrack
\,.
\label{equ:KirPioFirComCar}
\end{equation} 
In terms of \(\mathbf{P}\), quasi-static mechanical equilibrium takes the form 
\begin{equation}
\mathop{\mathrm{div}}\mathbf{P}
=\left\lbrack
\begin{array}{c}
P_{11,1}+P_{12,2}+P_{13,3}
\\
P_{21,1}+P_{22,2}+P_{23,3}
\\
P_{31,1}+P_{32,2}+P_{33,3}
\end{array}
\right\rbrack
=\bm{0}
\,,
\label{equ:EquMecStaQua}
\end{equation}
with \(P_{\!ij,k}:=\partial P_{\!ij}/\partial x_{k}\). 

\begin{figure}[H]
\begin{subfigure}[t]{\textwidth}
\includegraphics[width=0.86\textwidth]{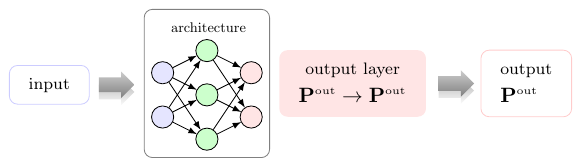}
\caption{Physics-guided or -informed NA for stress.}
\vspace{3mm}
\end{subfigure}
\begin{subfigure}[t]{\textwidth}
\includegraphics[width=0.99\textwidth]{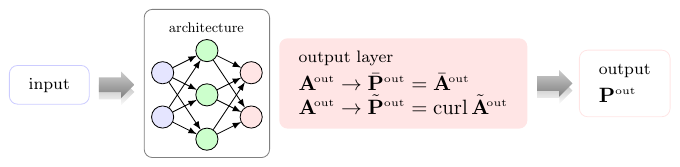}
\caption{Physics-encoded NA for divergence-free stress based on stress potential 
\(\mathbf{A}\) and split \(\mathbf{P}=\bar{\mathbf{P}}+\tilde{\mathbf{P}}\). 
See text for explanation. 
} 
\end{subfigure}
\caption{Schematics of physics-constrained NAs. 
In all cases, input (e.g., material properties, boundary conditions,  
or the deformation history) is transferred by the input layer (violet) 
to the hidden layers (green; only one is shown for simplicity). These 
in turn transform the input into \(\mathbf{P}^{\textrm{out}}\) 
via an output layer (light red).  See text for more details and discussion.} 
\label{fig:ArcANcP}
\end{figure}

As usual, the training of NAs is based on data; in the current case, 
the relevant data 
\(\mathbf{P}_{\!1}^{\textrm{dat}}, 
\ldots,
\mathbf{P}_{\!\smash{n_{\mathrm{dat}}}}^{\textrm{dat}}\) 
for the quasi-static equilibrium stress field satisfy 
\(\mathop{\mathrm{div}}\mathbf{P}_{\!\smash{a}}^{\textrm{dat}}=\bm{0}\) 
for \(a=1,\ldots,n_{\mathrm{dat}}\). 
In principle, then, the output \(\mathbf{P}^{\textrm{out}}\) 
of any NA trained with such data should also 
satisfy \(\mathop{\mathrm{div}}\mathbf{P}^{\textrm{out}}
=\bm{0}\); in practice of course, 
\(\mathop{\mathrm{div}}\mathbf{P}^{\textrm{out}}\neq\bm{0}\), 
due for example to issues such as the inherent sparsity of the data. 
As shown in more detail below, encoding this constraint in the NA architecture 
is the most robust way to ensure that \(\mathbf{P}^{\textrm{out}}\) is 
divergence-free. In the current work, this encoding is based on a stress 
potential \(\mathbf{A}\). A schematic comparison of this with 
other physics-constrained NAs for the stress field is given in \Cref{fig:ArcANcP}. 
The hidden layers (green in \Cref{fig:ArcANcP}) in NNs are 
based for example on a convolutional architecture such as U-Net 
\cite[e.g.,][]{mianroodi2021teaching,khorrami2023artificial} or other 
"deep" architectures \cite[e.g.,][]{Faroughi2024}. In the FNO, 
these layers are parameterized by linear weights and biases as well 
as by the kernel of a linear integral operator \cite[e.g.,][]{Li2021}. 
As depicted in \Cref{fig:ArcANcP}(b), the output layer (light red) in 
the physics-encoded case projects the potential field 
\(\mathbf{A}^{\!\textrm{out}}\) from the hidden layers onto its mean value 
\(\bar{\mathbf{A}}^{\!\textrm{out}}\) and its fluctuation part 
\(\tilde{\mathbf{A}}^{\!\textrm{out}}
=\mathbf{A}^{\!\textrm{out}}-\bar{\mathbf{A}}^{\!\textrm{out}}\) 
on the region of interest 
%\color{red}
(see the Supplementary Information). 
\color{black} 
These in turn are transformed by the output layer into 
\(\bar{\mathbf{P}}^{\textrm{out}}\) 
and \(\tilde{\mathbf{P}}^{\textrm{out}}\), respectively, such that 
\(\mathop{\mathrm{div}}\mathbf{P}^{\textrm{out}}
=\mathop{\mathrm{div}}\tilde{\mathbf{P}}^{\textrm{out}}
=\mathop{\mathrm{div}}\mathop{\mathrm{curl}}
\tilde{\mathbf{A}}^{\!\textrm{out}}=\bm{0}\). 
In PeNNs, such operator relations are evaluated in discretized form 
for example via automatic differentiation given a differentiable 
architecture \cite[e.g.,][]{RichterPowell2022}. 
%\color{red} 
In the PeFNO developed in the current work, these are evaluated 
with the help of Fourier methods and the corresponding discretized 
algebraic relations in Fourier space as documented in the 
Supplementary Information. 
\color{black} 

\section{Example application}
\label{sec:ExaCom}

As a computational example of the current PeFNO for divergence-free 
stress, consider its application to the surrogate modeling of the stress 
field in a heterogeneous solid with a grain microstructure. 
For comparison, analogous Pg- and PiFNOs are also considered. 
%\color{red}
Since the main purpose of the paper is the development of methods for 
material modeling, the computational example is based for simplicity on 
a number of simplifying assumptions. 
These include (i) isotropic elastic grains, 
(ii) limitation of FNO optimization to training, and related to this 
(iii) use of fixed values for FNO hyperparameters. 
Some of these simplifications are discussed in more detail below. 
\color{black}

\subsection{Synthetic stress field data for training}

Stress field data for FNO training are obtained from the numerical solution 
of a BVP for quasi-static mechanical equilibrium on a periodic unit cell \(U\) 
containing the grain microstructure. To this end, the stress in each grain is 
modeled here by the isotropic elastic Saint Venant-Kirchhoff relation 
\begin{equation}
\mathbf{P}
=\lambda\,(\mathbf{I}\cdot\mathbf{E})\,\mathbf{F}
+2\mu\,\mathbf{F}\mathbf{E}
\label{equ:ElaKirVenSai}
\end{equation} 
depending on the Lam\'e elastic moduli \(\lambda\) and \(\mu\). 
In this relation, \(\mathbf{F}\) and 
\begin{equation}
\mathbf{E}=\tfrac{1}{2}(\mathbf{F}^{\mathrm{T}}\mathbf{F}-\mathbf{I})
\label{equ:StnGre}
\end{equation} 
represent the matrices of Cartesian components 
of the deformation gradient and the symmetric Green strain, respectively, 
analogous to \Cref{equ:KirPioFirComCar} for \(\mathbf{P}\). 
In addition, \(\mathbf{I}\cdot\mathbf{E}=E_{11}+E_{22}+E_{33}\) 
is the trace of \(\mathbf{E}\). 

\begin{figure}[H]
\centering
\begin{subfigure}[t]{0.35\textwidth}
\includegraphics[width=\textwidth]{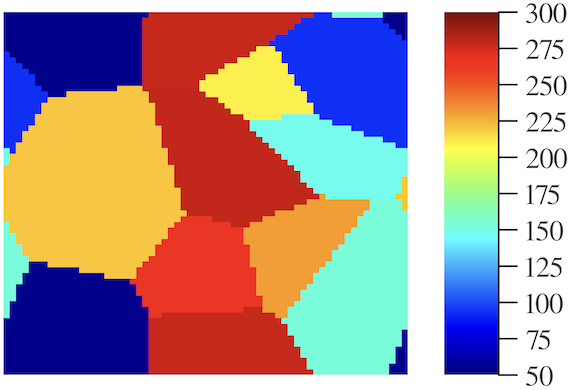}
\caption{\hbox{\(E\) [GPa]}}
\end{subfigure}
\hspace{10mm}
\begin{subfigure}[t]{0.35\textwidth}
\includegraphics[width=\textwidth]{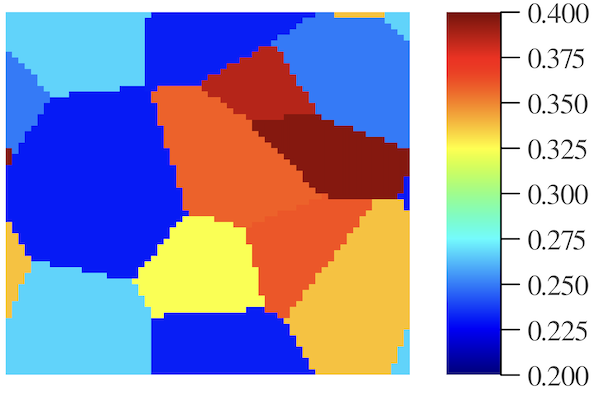}
\caption{\hbox{\(\nu\)}}
\end{subfigure}
\caption{Example microstructure in 2D unit cell \(U\) 
(\(x_{1}\) horizontal, \(x_{2}\) vertical).} 
\label{fig:ProMatDat}
\end{figure}

Each microstructure 
in the data set represents a random distribution of grain morphologies 
and grain material property value distributions for 
Young's modulus \(E\in\lbrack 50,300\rbrack\) GPa and Poisson's ratio 
\(\nu\in\lbrack 0.2,0.4\rbrack\) determining 
\(\lambda=E\nu/((1+\nu)(1-2\nu))\) and \(\mu=E/(2(1+\nu))\) 
in \Cref{equ:ElaKirVenSai}. An example of such a microstructure is 
shown in \Cref{fig:ProMatDat}.
The stress data set is generated for a set of \(n_{\mathrm{dat}}\) 
such microstructures, all subject to the same prescribed mean local 
deformation \(\bar{\mathbf{F}}\). 
The corresponding BVP based in particular 
on Equations \eqref{equ:EquMecStaQua}-\eqref{equ:ElaKirVenSai} 
has been implemented and numerically solved using spectral methods 
\citep{willot2015fourier,khorrami2020development} 
and the software toolkit DAMASK \citep{roters2019damask}. 

\subsection{FNO training} 

Training of the PgFNO and PeFNO is based on the weighted mean absolute error 
\begin{equation}
\textstyle
L_{\mathrm{dat}}
:=\dfrac{1}{n_{\mathrm{dat}}}
\sum_{a=1}^{n_{\mathrm{dat}}}
\sum_{b=1}^{n_{\mathrm{dis}}}
|\mathbf{W}_{\!ab}^{\mathrm{dat}}
\cdot
(\mathbf{P}_{\!ab}^{\mathrm{out}}-\mathbf{P}_{\!ab}^{\mathrm{dat}})|
\label{equ:FunLosEncGuiPhy}
\end{equation}
for the loss function, where  
\(n_{\mathrm{dis}}\) is the number of discretization points. 
In this relation, \(\mathbf{P}_{\!\smash{ab}}\) is the 
stress field value at the discretization point position 
\(\mathbf{x}_{b}=(x_{1},x_{2},x_{3})_{b}\) 
(\(b=1,\ldots,n_{\mathrm{dis}}\)) in the \(a^{\mathrm{th}}\) 
microstructure (\(a=1,\ldots,n_{\mathrm{dat}}\)) of the data set,  
and \(\mathbf{W}_{\!ab}^{\mathrm{dat}}\) the corresponding weighting 
matrix. Training is based on \(n_{\mathrm{dat}}=1000\) in the sequel.

For the current example based on the grain microstructure in 
\Cref{fig:ProMatDat}, values for the components of 
\(\mathbf{W}_{\!ab}^{\mathrm{dat}}\) are determined as follows. 
Given a uniform numerical discretization of \(U\), the number of data 
in regions of uniform stress (i.e., grain interiors) is much larger than 
the number of data in regions of non-uniform stress (i.e., grain boundaries). 
Consequently, unweighted training automatically results in 
trained FNO stress field output which underestimates spatial variations 
in the stress field, especially near grain boundaries. 
To compensate for this bias, \(\mathbf{W}_{\!ab}^{\mathrm{dat}}\) 
components are proportional to the sum of 1 and 
a term determined by the magnitude of the gradient of the 
corresponding component of \(\mathbf{P}_{\!a}^{\textrm{dat}}\) 
at \(\mathbf{x}_{b}\) normalized by this component 
(examples of the components of \(\mathbf{W}_{\!ab}^{\mathrm{dat}}\) 
are shown in \Cref{fig:WeiFit} below). 

For the PiFNO, the loss function takes the form
\begin{equation}
\textstyle
L:=L_{\textrm{dat}}+c\,L_{\textrm{div}}
\,,\quad
L_{\textrm{div}}
:=\dfrac{1}{n_{\mathrm{dat}}}
\sum_{a=1}^{n_{\mathrm{dat}}}
\sum_{b=1}^{n_{\mathrm{dis}}}
|(\mathop{\mathrm{div}}\mathbf{P}_{\!a}^{\smash{\mathrm{out}}})_{b}|
\,,
\label{equ:FunLosInfPhy}
\end{equation} 
where \(c\) is a scalar hyperparameter. In contrast to \(L_{\mathrm{dat}}\), 
note that \(L_{\mathrm{div}}\) is unweighted. 
Like the operator relations in the output layer of the PeFNO 
based on \Cref{fig:ArcANcP}(b), 
\((\mathop{\mathrm{div}}\mathbf{P}_{\!a}^{\smash{\mathrm{out}}})_{b}\) 
and \(L_{\textrm{div}}\) are evaluated with the help of 
Fourier methods and the corresponding algebraic relations in Fourier space 
%\color{red}
(given in the Supplementary Information). 
\color{black}

\subsection{FNO architecture and hyperparameters} 

Following in essence \cite{Li2021}, the architecture of 
all FNOs in what follows consists of 1 input layer 
(violet in \Cref{fig:ArcANcP}), 
4 hidden layers (light green in \Cref{fig:ArcANcP}), 
and 1 output layer (light red in \Cref{fig:ArcANcP}). Each 
hidden layers contains 20 neurons or "channels" 
(only three are shown in \Cref{fig:ArcANcP} for simplicity). 

The number of neurons per hidden layer, 
or the number of hidden layers in the architecture, represent so-called 
hyperparameters. In principle, their values should be optimized during 
testing and validation of the trained FNO. 
For simplicity, however, we follow \cite{Li2021} in working with a fixed 
value for this. They established this value empirically, and showed that 
increasing this number beyond 20 led to no significant increase in 
robustness of the trained FNO. Likewise for simplicity, the value of 
the scalar hyperparameter \(c\) in \Cref{equ:FunLosInfPhy} is also 
fixed here. Unless otherwise specified, \(c=10^{-1}\) in the sequel. 

\subsection{Results for the stress field} 

\begin{figure}[H]
\centering
\begin{subfigure}[t]{0.35\textwidth}
\includegraphics[width=\textwidth]{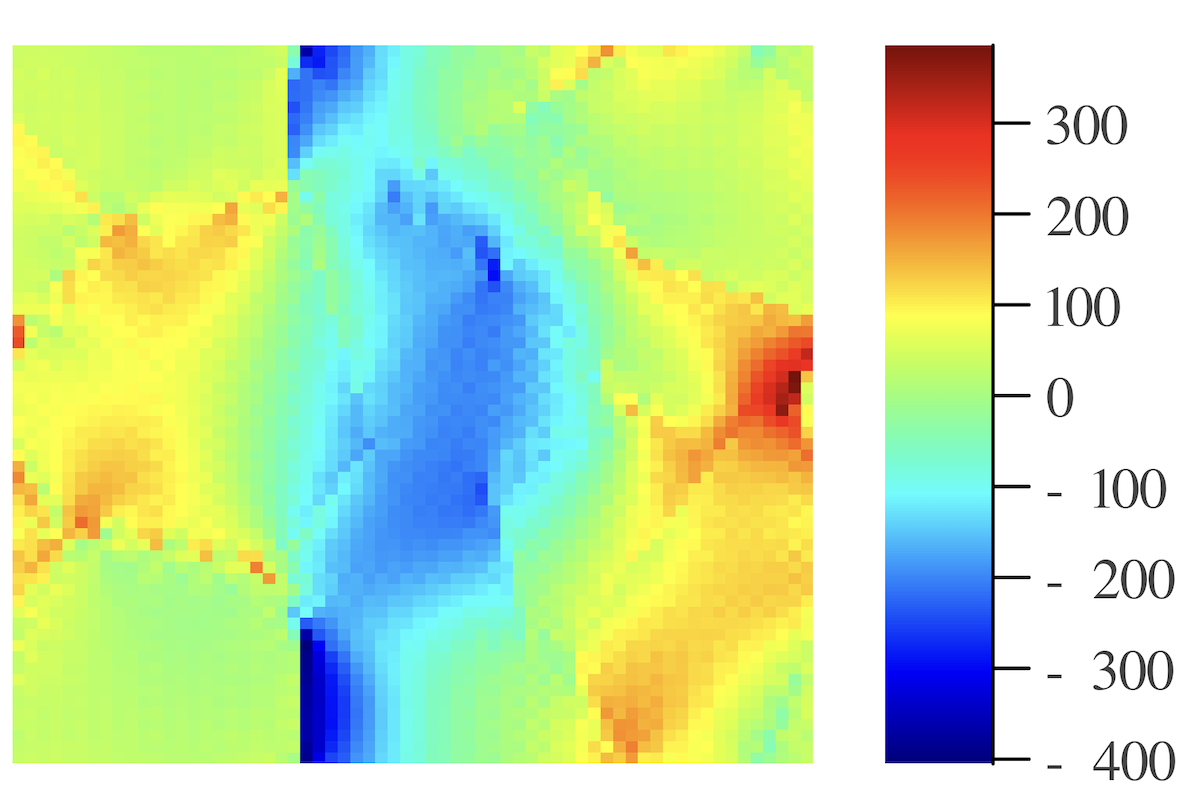}
\caption{\hbox{\(P_{\!11}^{\mathrm{dat}}\)}}
\vspace{3mm}
\end{subfigure}
\hspace{10mm}
\begin{subfigure}[t]{0.35\textwidth}
\includegraphics[width=\textwidth]{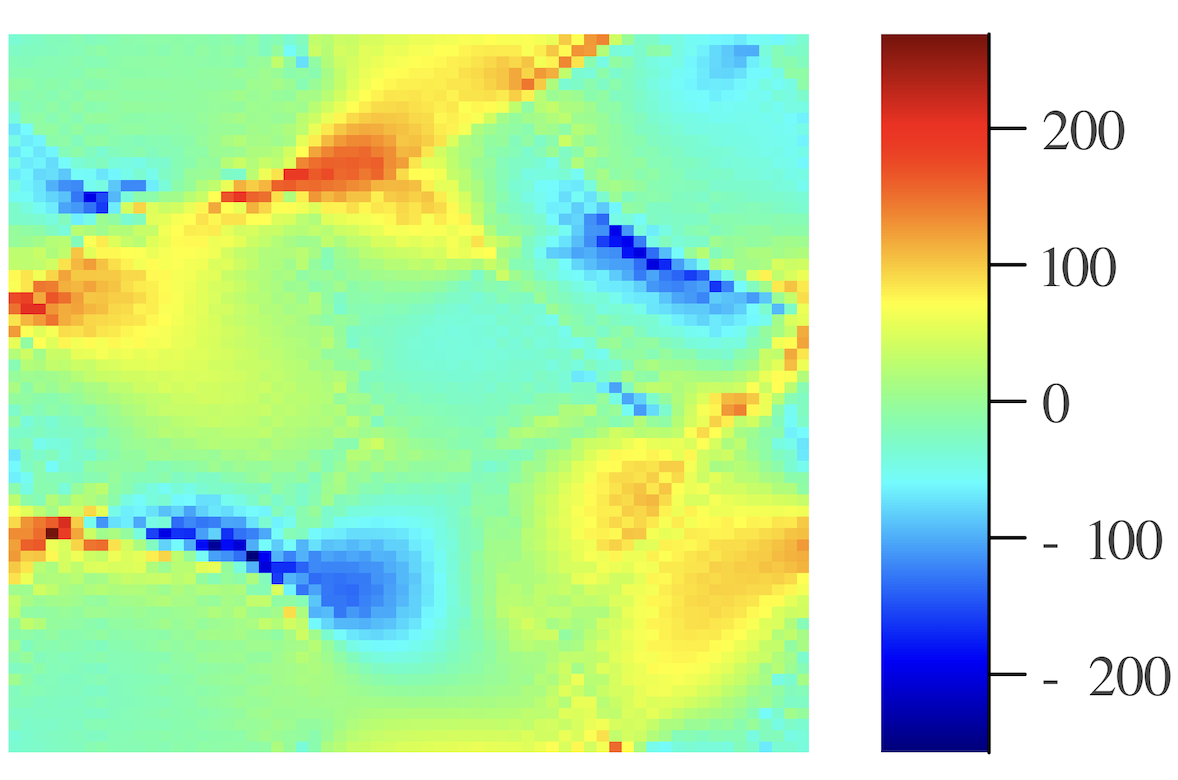}
\caption{\hbox{\(P_{\!12}^{\mathrm{dat}}\)}}
\vspace{3mm}
\end{subfigure}
\begin{subfigure}[t]{0.35\textwidth}
\includegraphics[width=\textwidth]{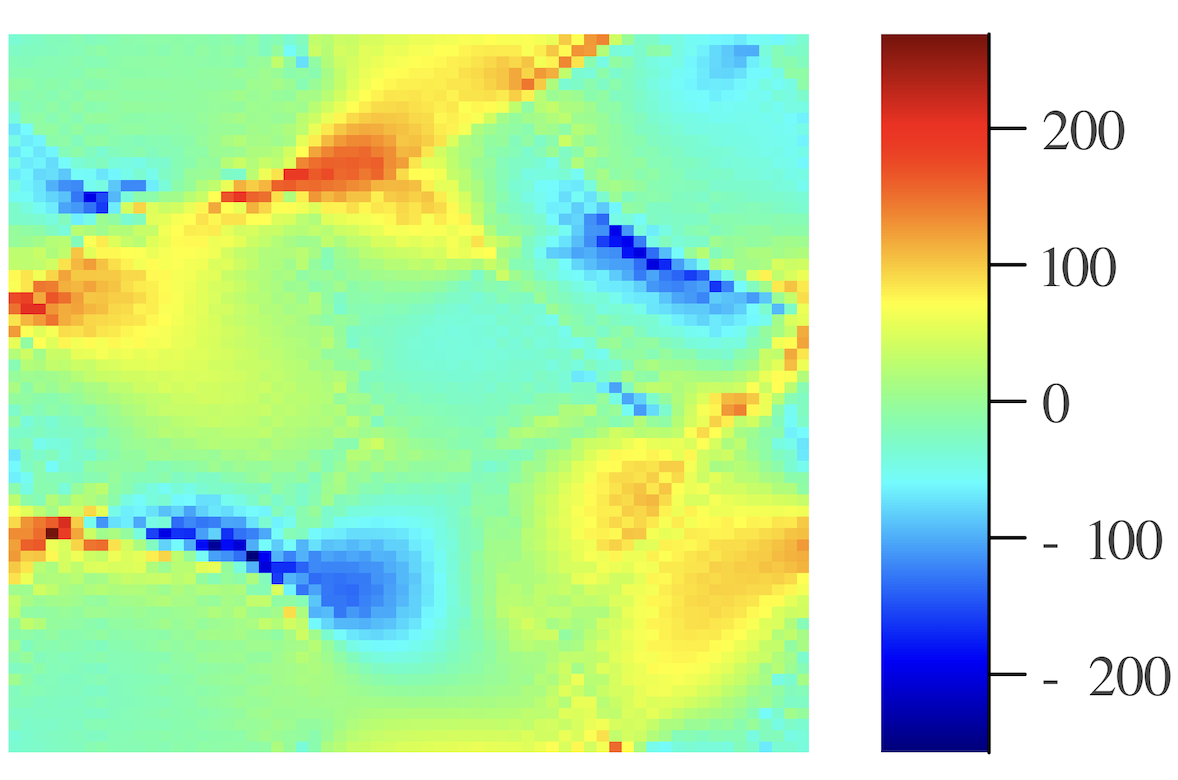}
\caption{\hbox{\(P_{\!21}^{\mathrm{dat}}\)}}
\end{subfigure}
\hspace{10mm}
\begin{subfigure}[t]{0.35\textwidth}
\includegraphics[width=\textwidth]{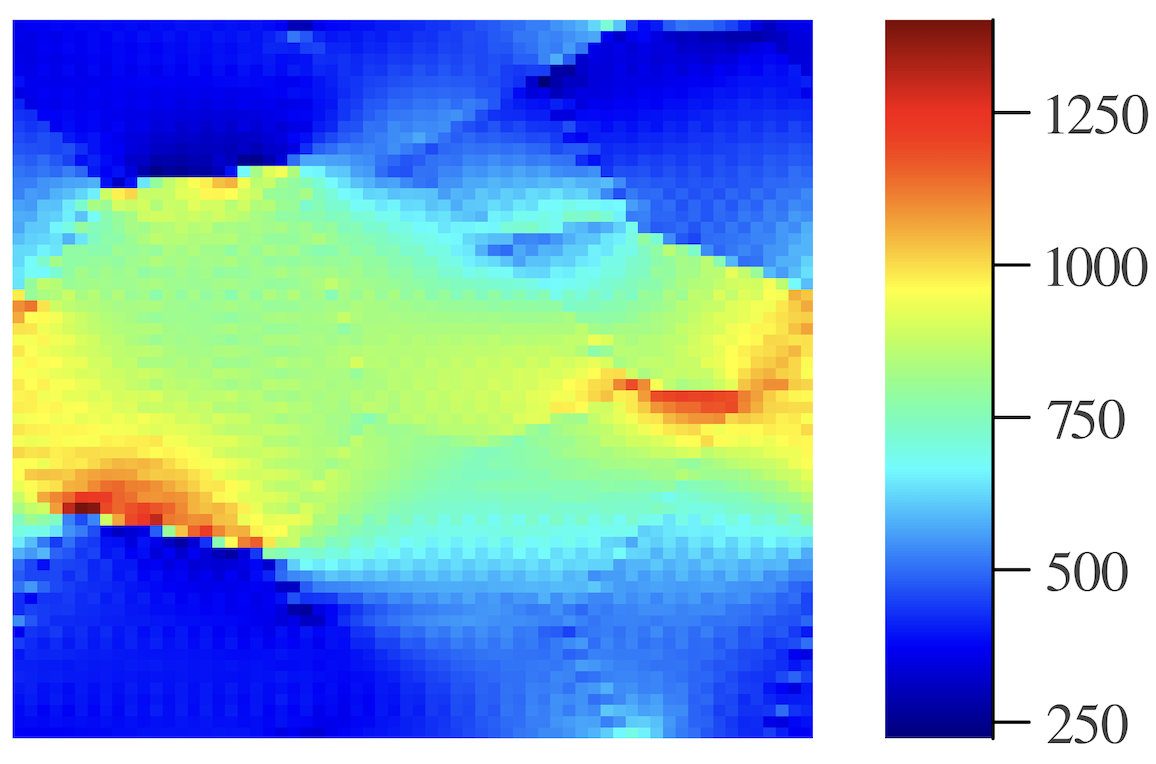}
\caption{\hbox{\(P_{\!22}^{\mathrm{dat}}\)}}
\end{subfigure}
\caption{Stress field data in \(U\) 
for uniaxial extension of the microstructure in \Cref{fig:ProMatDat} 
in the \(x_{2}\) direction. 
All values in MPa. 
Note the differences in scale. 
} 
\label{fig:DatStsComCar}
\end{figure}

Restricting attention for simplicity to plane deformation of a 
square unit cell \(U\) of side-length \(l\) in the \(x_{1}\) (horizontal), 
\(x_{2}\) (vertical) plane, note that \(F_{33}=1\) is constant and 
\(F_{13}\), \(F_{23}\), \(F_{31}\), \(F_{32}\) are identically zero. Then 
\(P_{13}=P_{23}=0\) and \(P_{31}=P_{32}=0\) hold identically via 
\Cref{equ:ElaKirVenSai}. In the context of plane deformation, the unit cell 
containing the microstructure is subject to the mean deformation gradient 
\(\bar{F}_{11}=1\), 
\(\bar{F}_{22}=1.004\), 
\(\bar{F}_{33}=1\) (all other \(\bar{F}_{ij}=0\)) 
corresponding to uniaxial extension in the \(x_{2}\) direction and 
a mean Green strain of \(\bar{E}_{22}=0.401\%\) (all other \(\bar{E}_{ij}=0\)). 
%\color{red}
The data and trained FNO results presented in the following are based on 
the material property distribution shown in \Cref{fig:ProMatDat} 
(not a part of the training data set) and a uniform 
discretization of \(U\) based on a square lattice or 2D grid with numerical 
resolution of \(n_{\mathrm{dis}}=64\) in each dimension. 
\color{black}
The corresponding stress field data are shown in \Cref{fig:DatStsComCar}. 
Since \(P_{\!12}^{\mathrm{dat}}\) is quite similar to 
\(P_{\!21}^{\mathrm{dat}}\) (due in particular to isotropy), 
attention is focused for simplicity on \(P_{11}\), \(P_{21}\), and \(P_{22}\) 
in what follows. 

\begin{figure}[H]
\centering
\begin{subfigure}[t]{\textwidth}
\centering
\includegraphics[width=0.35\textwidth]{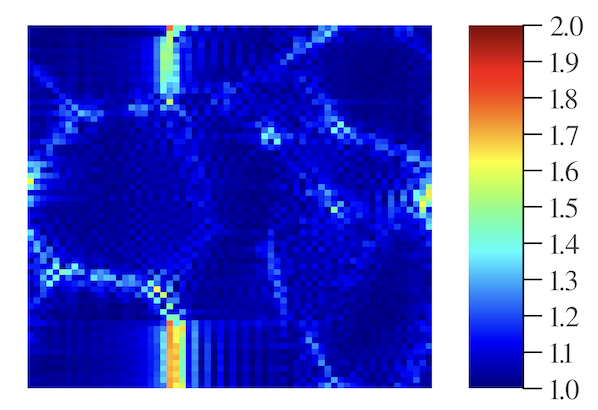}
\caption{\hbox{\(W_{\!11}^{\mathrm{dat}}\)}}
\vspace{3mm}
\end{subfigure}
\begin{subfigure}[t]{0.35\textwidth}
\includegraphics[width=\textwidth]{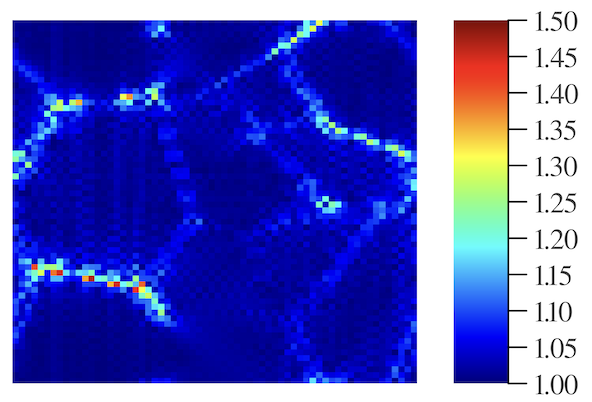}
\caption{\hbox{\(W_{\!21}^{\mathrm{dat}}\)}}
\end{subfigure}
\hspace{10mm}
\begin{subfigure}[t]{0.35\textwidth}
\includegraphics[width=\textwidth]{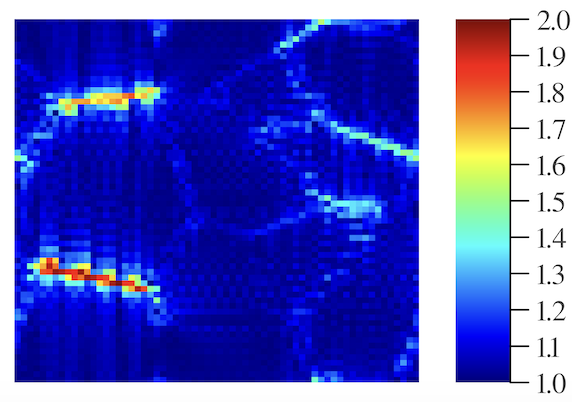}
\caption{\hbox{\(W_{\!22}^{\mathrm{dat}}\)}}
\end{subfigure}
\caption{Example components of \(\mathbf{W}^{\mathrm{dat}}\) 
[1/\(l\)] for data in \Cref{fig:DatStsComCar}. 
}
\label{fig:WeiFit}
\end{figure}

Selected components of \(\mathbf{W}^{\mathrm{dat}}\) for the data in 
\Cref{fig:DatStsComCar} are shown in \Cref{fig:WeiFit}. 
Since the components of \(\mathbf{W}^{\mathrm{dat}}\) 
are determined by the magnitude of the gradient of the 
corresponding component of \(\mathbf{P}_{\!a}^{\textrm{dat}}\), 
the results in \Cref{fig:WeiFit} show that the stress field gradients are 
largest at grain boundaries. This is also implied by direct inspection of 
\Cref{fig:DatStsComCar}. More specifically, this is the case at the 
grain boundaries with the largest material property contrast (in particular in \(E\)),  
as shown by comparison with the contrast in the 
material properties at grain boundaries in \Cref{fig:ProMatDat}. 

\begin{figure}[H]
\centering
\begin{subfigure}[t]{0.35\textwidth}
\includegraphics[width=\textwidth]{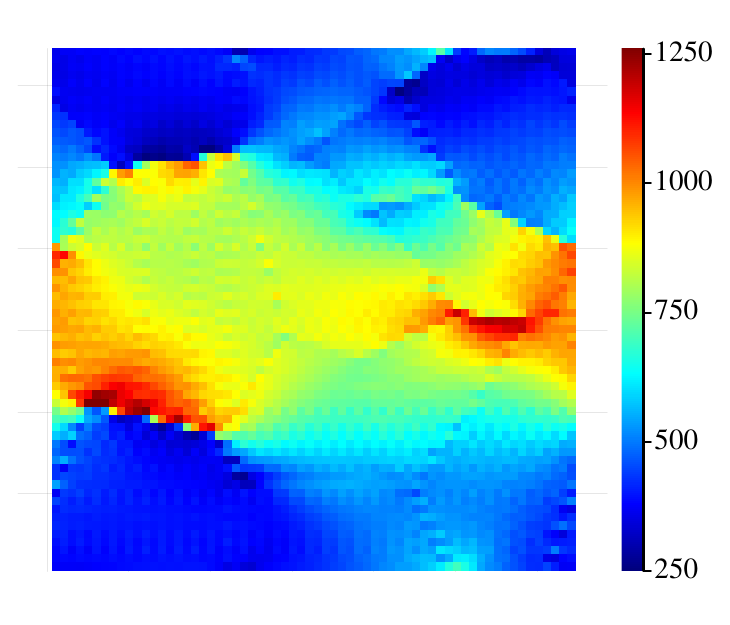}
\caption{\hbox{\(P_{\!22}^{\mathrm{dat}}\)}}
\vspace{3mm}
\end{subfigure}
\hspace{10mm}
\begin{subfigure}[t]{0.35\textwidth}
\includegraphics[width=\textwidth]{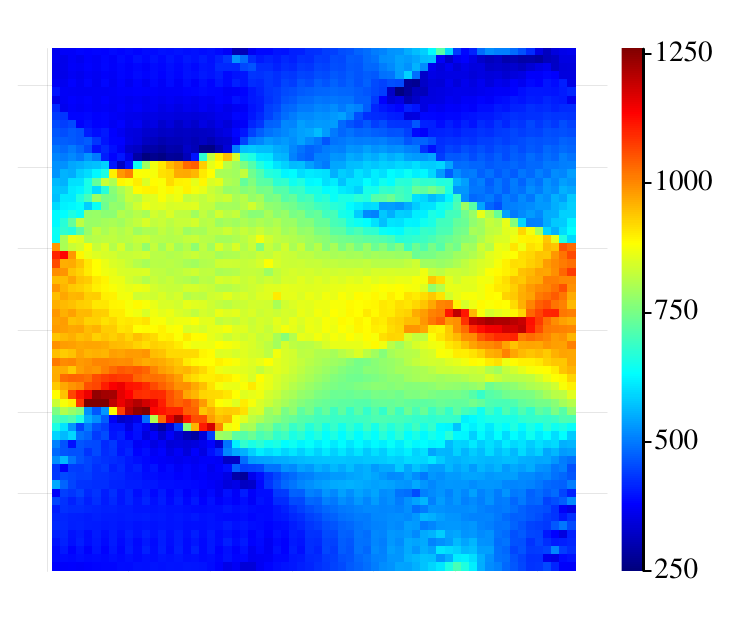}
\caption{\hbox{\(P_{\!22}^{\mathrm{out}}\) (PgFNO)}}
\vspace{3mm}
\end{subfigure}
\begin{subfigure}[t]{0.35\textwidth}
\includegraphics[width=\textwidth]{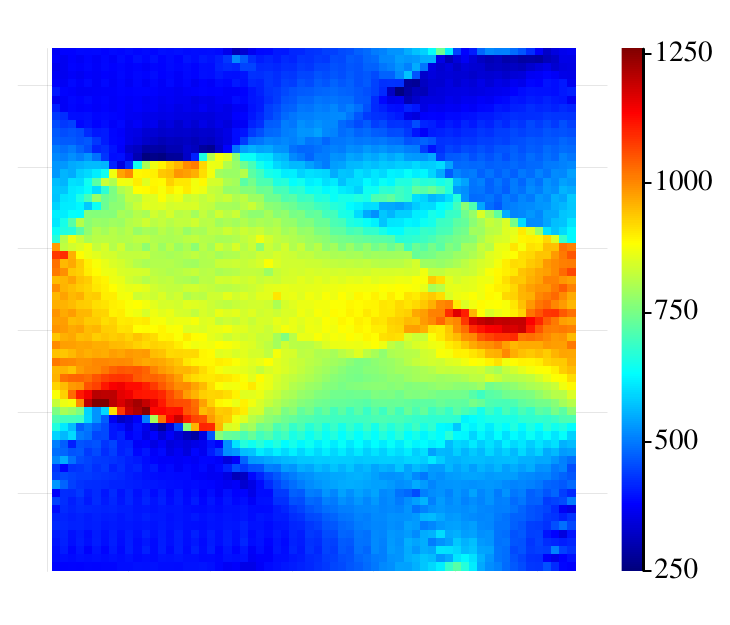}
\caption{\hbox{\(P_{\!22}^{\mathrm{out}}\) (PiFNO)}}
\end{subfigure}
\hspace{10mm}
\begin{subfigure}[t]{0.35\textwidth}
\includegraphics[width=\textwidth]{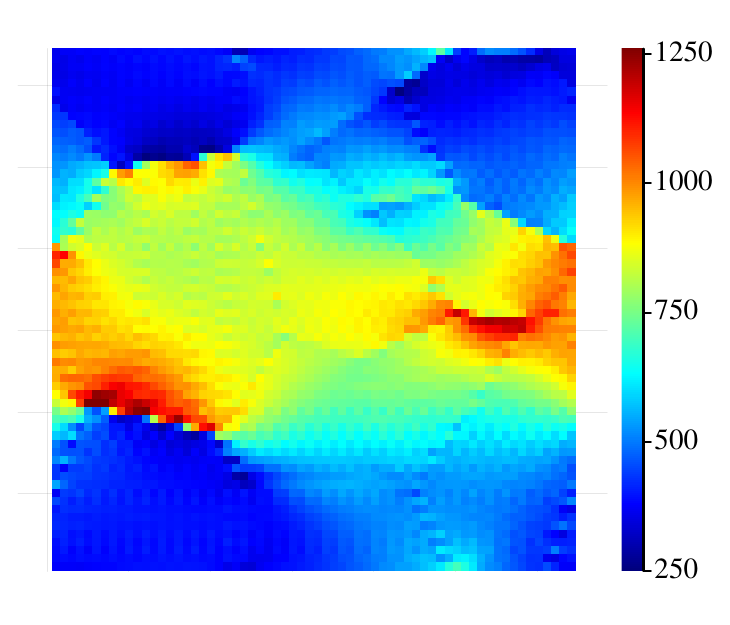}
\caption{\hbox{\(P_{\!22}^{\mathrm{out}}\) (PeFNO)}}
\end{subfigure}
\caption{Comparison of \(P_{\!22}^{\mathrm{dat}}\) and 
trained FNO results for \(P_{\!22}^{\mathrm{out}}\) in \(U\). 
All values in MPa. 
}
\label{fig:P33_comparison}
\end{figure}

\begin{figure}[H]
\centering
\begin{subfigure}[t]{0.35\textwidth}
\includegraphics[width=\textwidth]{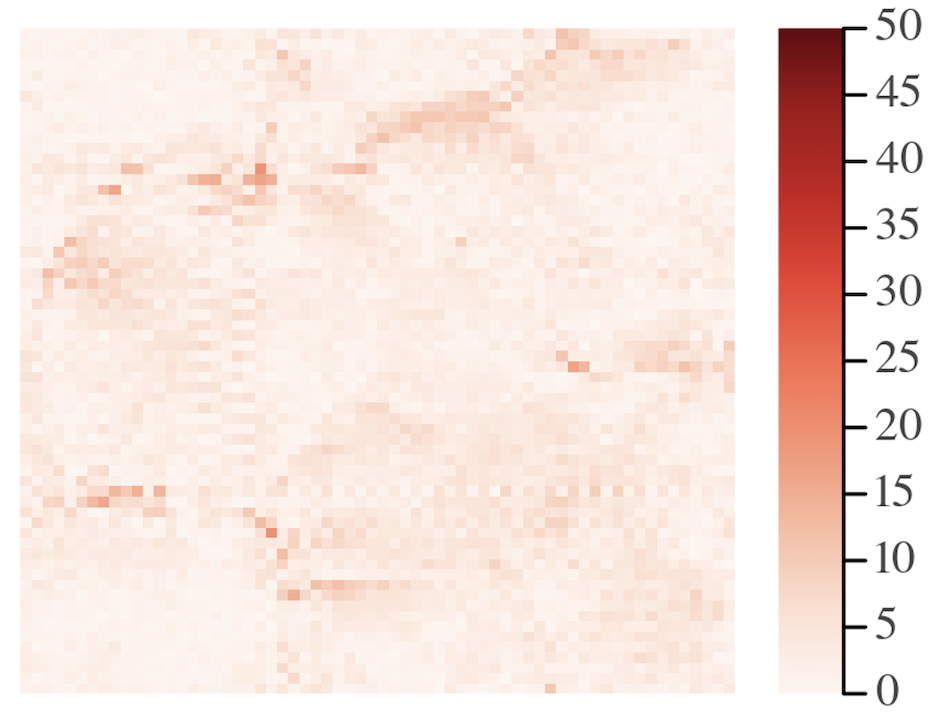}
\caption{\hbox{PgFNO}}
\vspace{3mm}
\end{subfigure}
\hspace{10mm}
\begin{subfigure}[t]{0.35\textwidth}
\includegraphics[width=\textwidth]{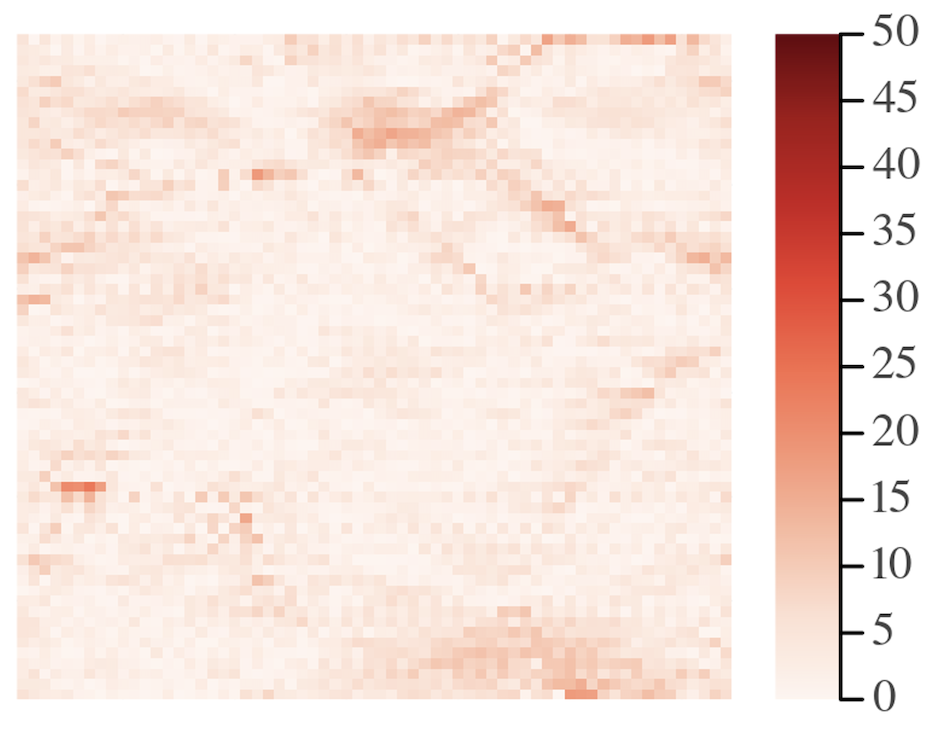}
\caption{\hbox{PeFNO}}
\vspace{3mm}
\end{subfigure}
\begin{subfigure}[t]{0.35\textwidth}
\includegraphics[width=\textwidth]{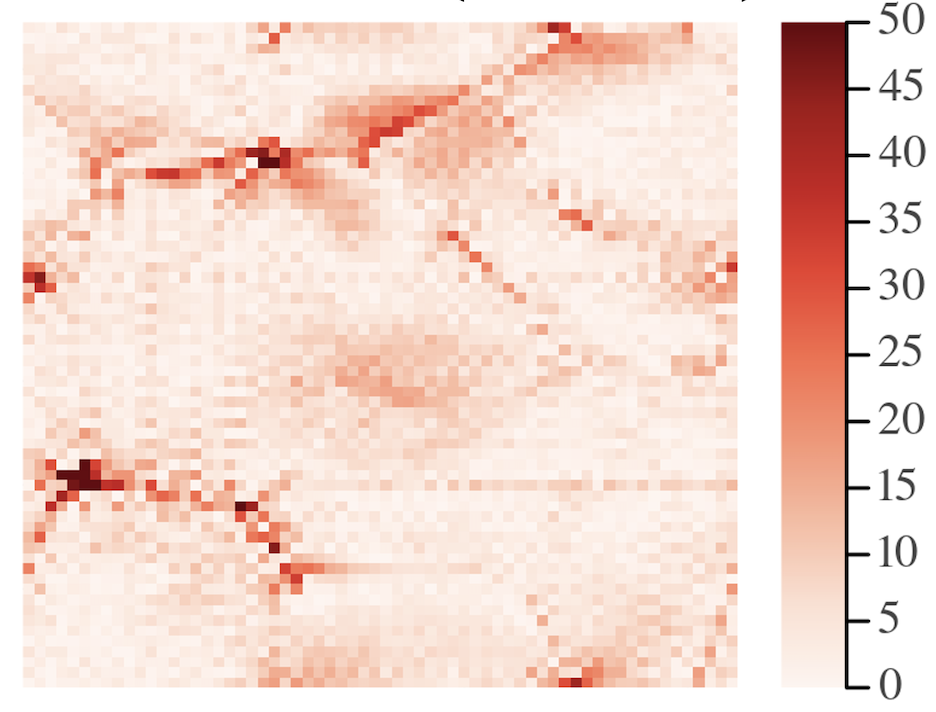}
\caption{\hbox{PiFNO, \(c=10^{-2}\)}}
\end{subfigure}
\hspace{10mm}
\begin{subfigure}[t]{0.35\textwidth}
\includegraphics[width=\textwidth]{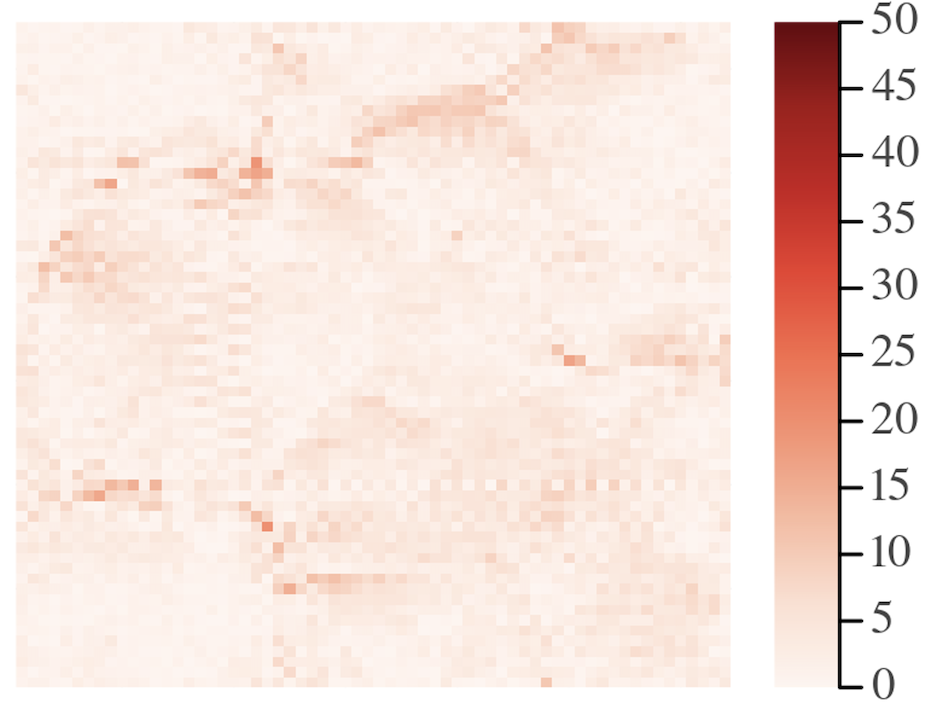}
\caption{\hbox{PiFNO, \(c=10^{-1}\)}}
\end{subfigure}
\caption{Magnitude 
\(|P_{22}^{\mathrm{out}}-P_{22}^{\mathrm{dat}}|\) of the error in 
\(P_{22}^{\mathrm{out}}\).
All values in MPa.
}
\label{fig:P33_dataError_comparison}
\end{figure}

\begin{figure}[H]
\centering
\begin{subfigure}[t]{0.35\textwidth}
\includegraphics[width=\textwidth]{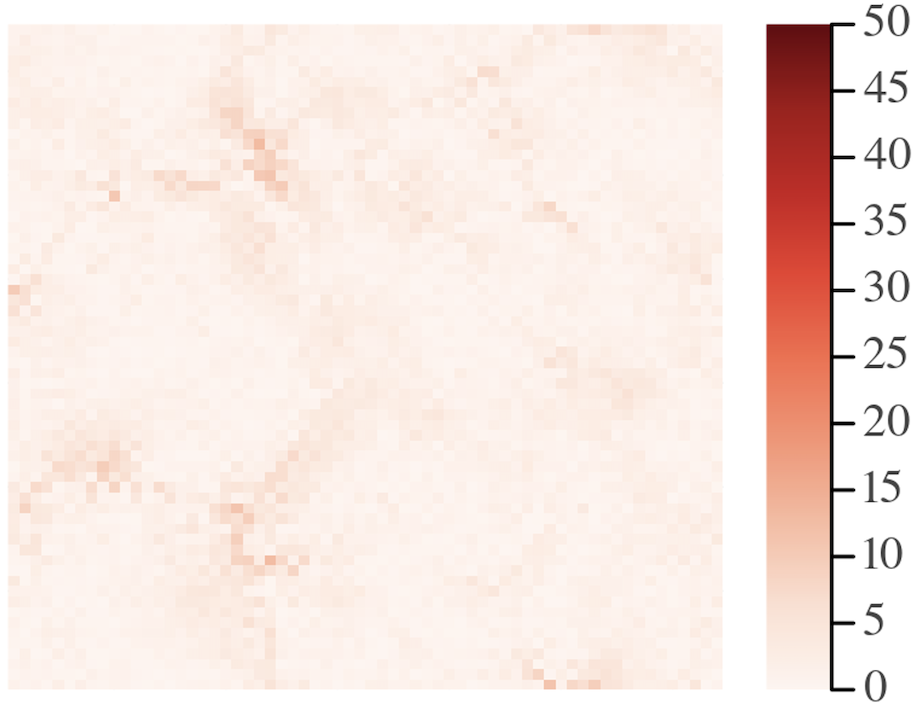}
\caption{\hbox{PgFNO}}
\vspace{3mm}
\end{subfigure}
\hspace{10mm}
\begin{subfigure}[t]{0.35\textwidth}
\includegraphics[width=\textwidth]{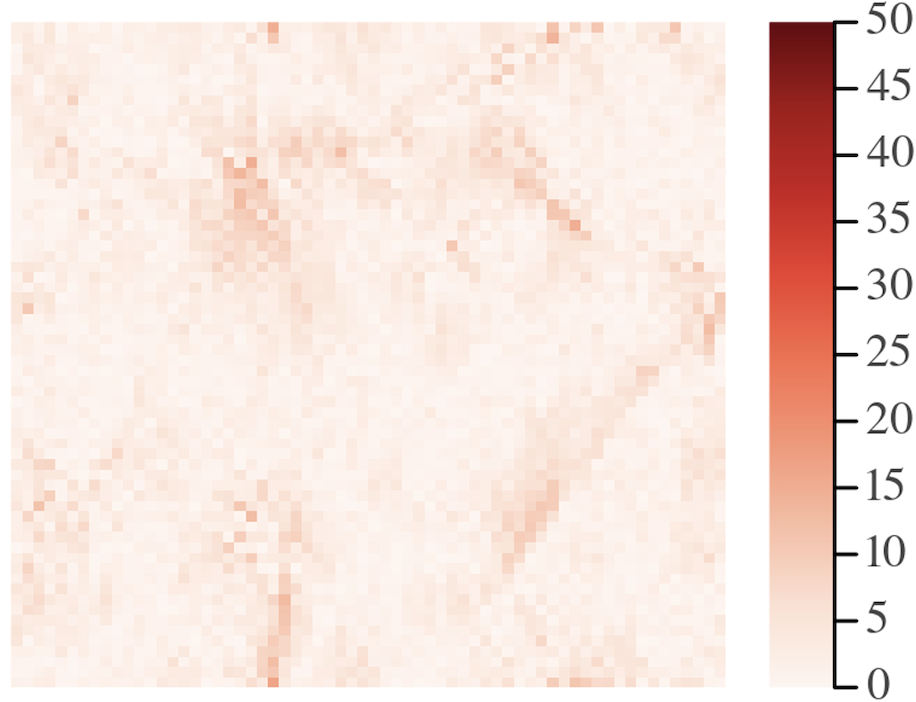}
\caption{\hbox{PeFNO}}
\vspace{3mm}
\end{subfigure}
\begin{subfigure}[t]{0.35\textwidth}
\centering
\includegraphics[width=\textwidth]{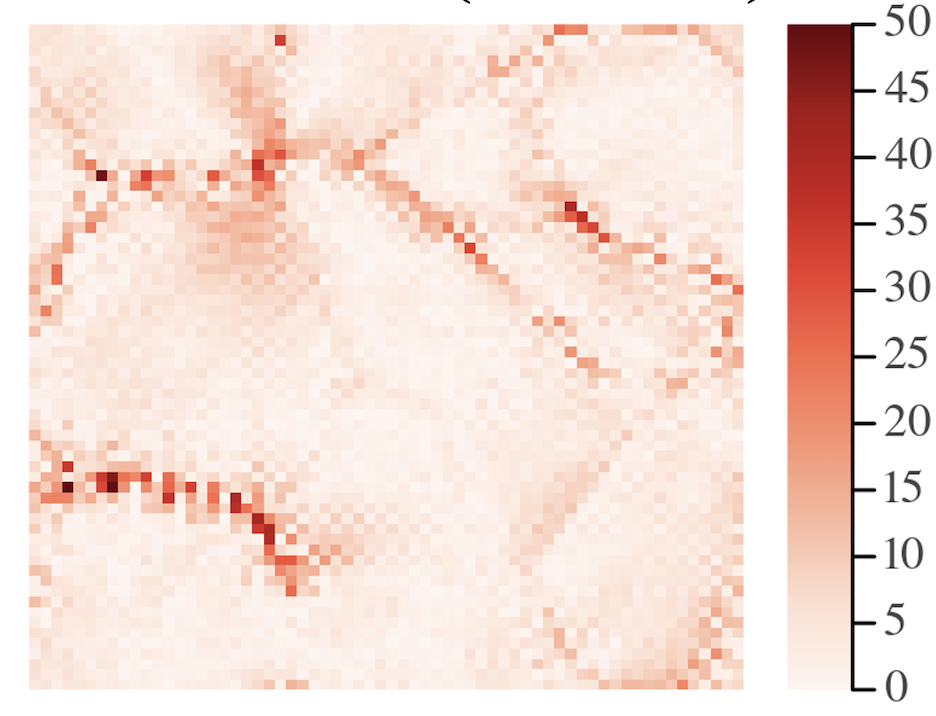}
\caption{\hbox{PiFNO}, \(c=10^{-2}\)}
\end{subfigure}
\hspace{10mm}
\begin{subfigure}[t]{0.35\textwidth}
\includegraphics[width=\textwidth]{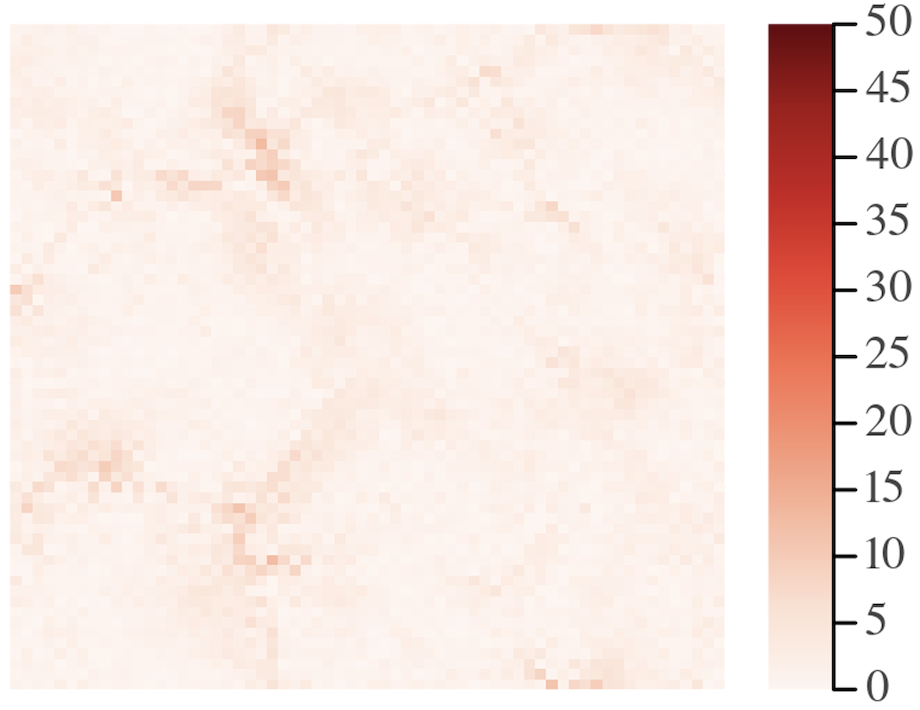}
\caption{\hbox{PiFNO}, \(c=10^{-1}\)}
\end{subfigure}
\caption{Magnitude 
\(|P_{21}^{\mathrm{out}}-P_{21}^{\mathrm{dat}}|\) 
of the error in \(P_{21}^{\mathrm{out}}\). 
All values in MPa.}
\label{fig:P32_dataError_comparison}
\end{figure}

\begin{figure}[H]
\centering
\begin{subfigure}[t]{0.35\textwidth}
\includegraphics[width=\textwidth]{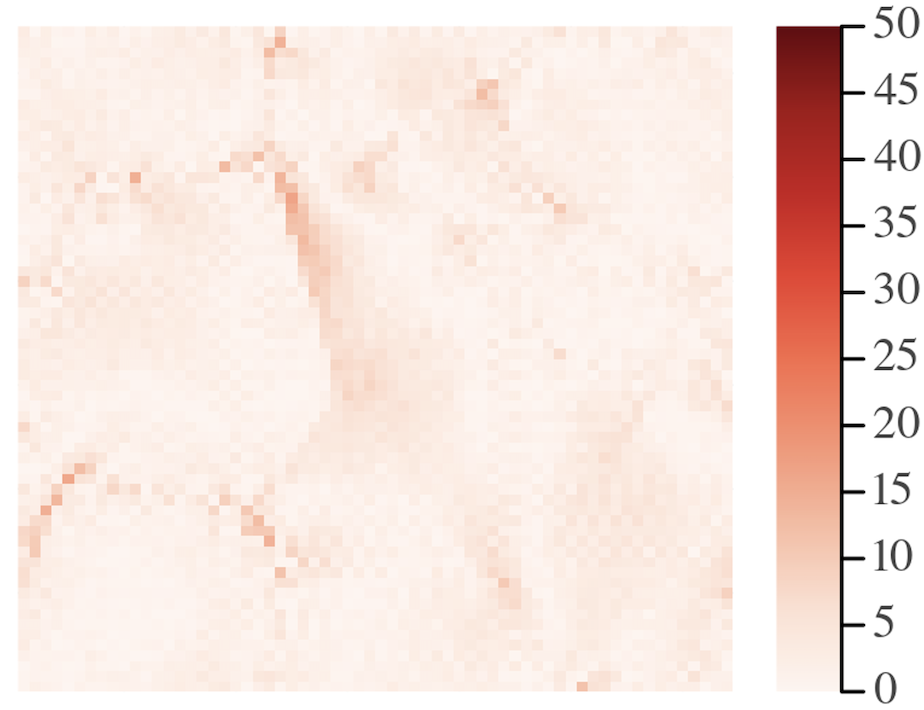}
\caption{\hbox{PgFNO}}
\vspace{3mm}
\end{subfigure}
\hspace{10mm}
\begin{subfigure}[t]{0.35\textwidth}
\includegraphics[width=\textwidth]{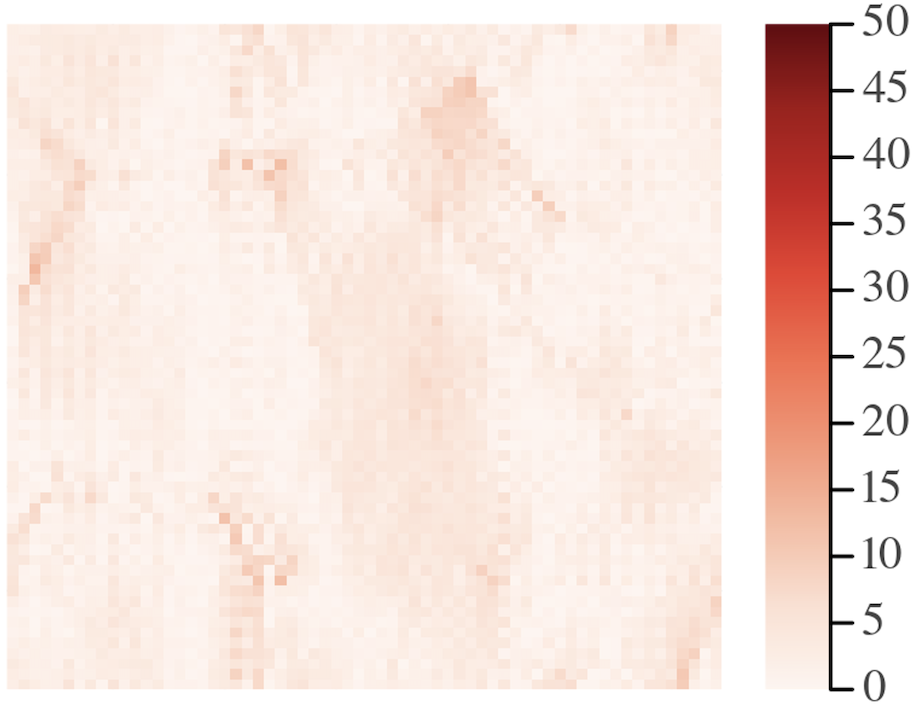}
\caption{\hbox{PeFNO}}
\vspace{3mm}
\end{subfigure}
\begin{subfigure}[t]{0.35\textwidth}
\includegraphics[width=\textwidth]{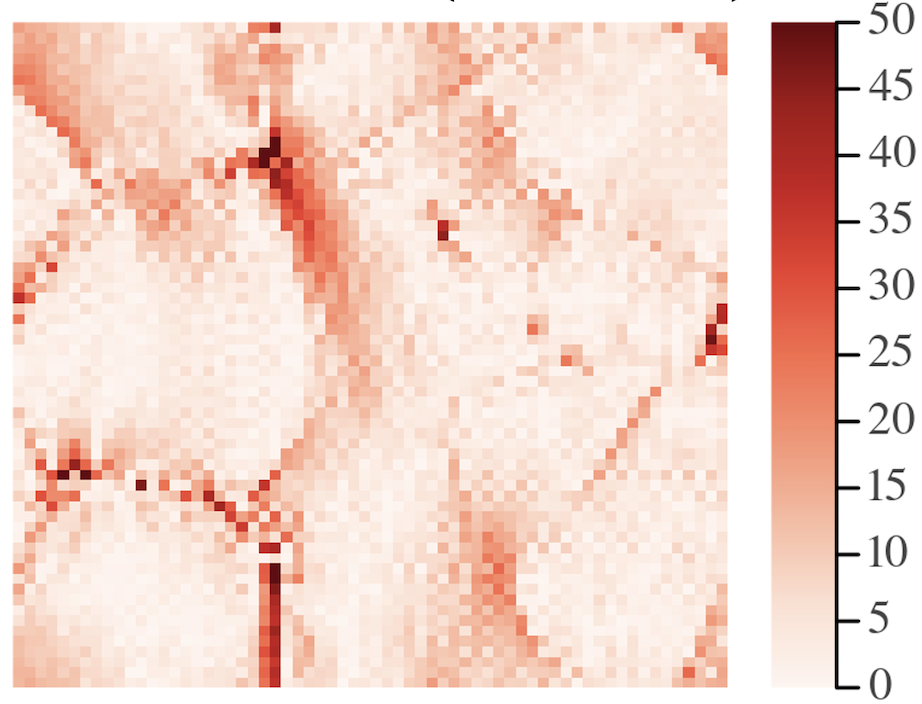}
\caption{\hbox{PiFNO, \(c=10^{-2}\)}}
\end{subfigure}
\hspace{10mm}
\begin{subfigure}[t]{0.35\textwidth}
\includegraphics[width=\textwidth]{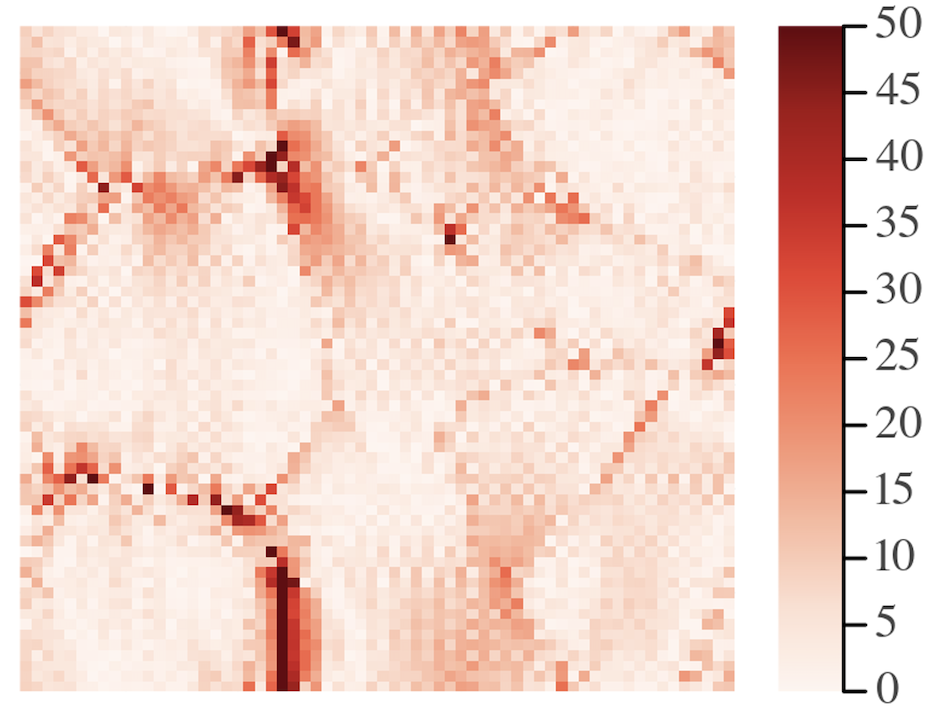}
\caption{\hbox{PiFNO, \(c=10^{-1}\)}}
\end{subfigure}
\caption{Magnitude \(|P_{11}^{\mathrm{out}}-P_{11}^{\mathrm{dat}}|\) 
of the error in \(P_{11}^{\mathrm{out}}\). 
All values in MPa.}
\label{fig:P22_dataError_comparison}
\end{figure}

Given that \(\bar{\mathbf{F}}\) is the same for all data, note that 
the material property distribution determined by \Cref{fig:ProMatDat} and 
the numerical discretization is the only input to the trained FNOs in what 
follows. Results for the largest stress component 
\(P_{22}\) are shown in \Cref{fig:P33_comparison}. 
The magnitude \(|P_{\!22}^{\mathrm{out}}-P_{\!22}^{\mathrm{dat}}|\) 
of the error in \(P_{\!22}^{\mathrm{out}}\) is displayed in 
\Cref{fig:P33_dataError_comparison}. 
Analogous error results for \(P_{21}^{\mathrm{out}}\) and 
\(P_{11}^{\mathrm{out}}\) are displayed in Figures 
\ref{fig:P32_dataError_comparison}-\ref{fig:P22_dataError_comparison}.
As evident in Figures 
\ref{fig:P33_dataError_comparison}-\ref{fig:P22_dataError_comparison}, 
\(P_{11}^{\mathrm{out}}\), 
\(P_{21}^{\mathrm{out}}\), and 
\(P_{22}^{\mathrm{out}}\) calculated by the trained PiFNO
exhibit the largest error of about 50 MPa for \(c=10^{-2}\). 
Comparison with the material 
property distribution in \Cref{fig:ProMatDat} implies that the magnitude 
of this error is correlated with the magnitude of the contrast in material 
properties across the corresponding grain boundary. Except for the error 
in \(P_{11}^{\mathrm{out}}\), increasing the value 
of \(c\) to \(10^{-1}\) results in a decrease of this error. 
Recall that training of both the PgFNO and PeFNO based 
on \Cref{equ:FunLosEncGuiPhy} is independent of 
\(L_{\textrm{div}}\). Consequently, 
the larger error in the output of the trained PiFNO
is clearly related to the inclusion of \(L_{\textrm{div}}\) 
in the loss function. Also, note that the weighting in 
\Cref{equ:FunLosEncGuiPhy} appears to be more effective 
at reducing error in the output of the trained PgFNO and PeFNO 
than in the trained PiFNO. A possible improvement 
here could be to include such weighting in \(L_{\textrm{div}}\) as well. 

\subsection{Results for the stress field divergence} 

\begin{figure}[H]
\centering
\begin{subfigure}[t]{0.35\textwidth}
\includegraphics[width=\textwidth]{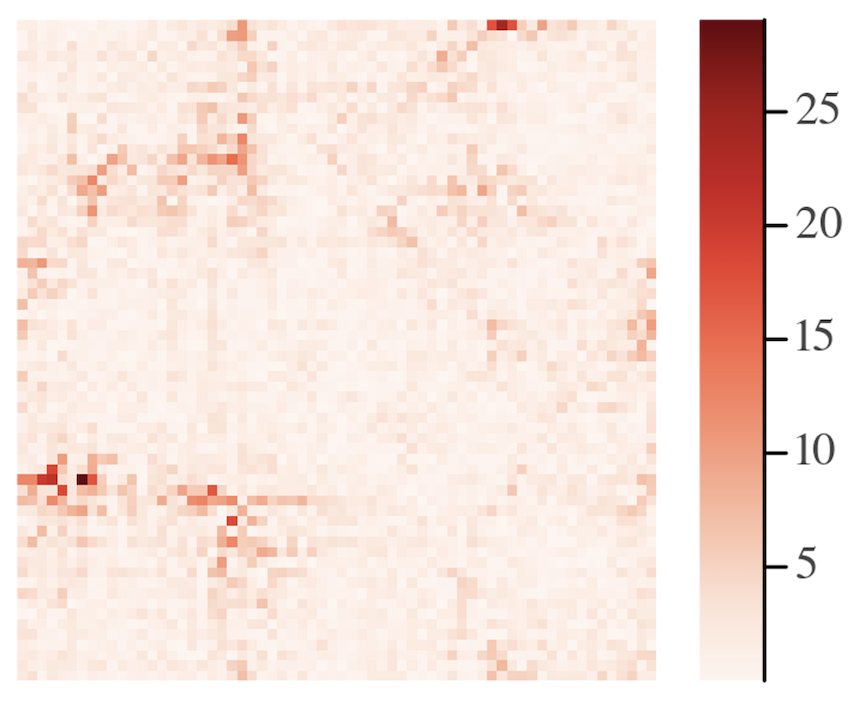}
\caption{\hbox{PgFNO}}
\vspace{3mm}
\end{subfigure}
\hspace{6mm}
\begin{subfigure}[t]{0.4\textwidth}
\includegraphics[width=\textwidth]{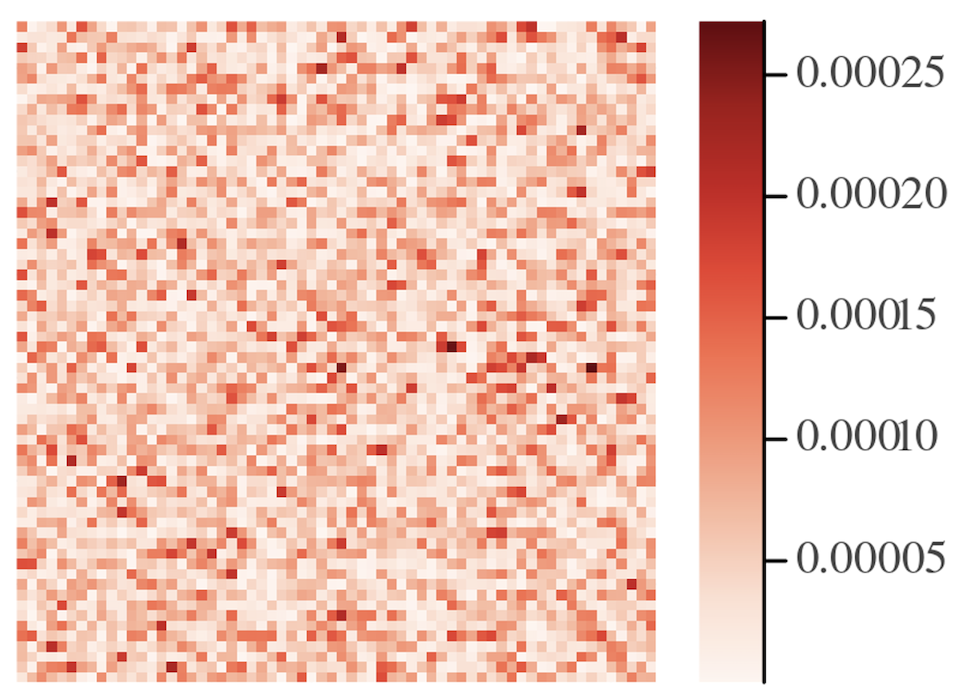}
\caption{\hbox{PeFNO}}
\vspace{3mm}
\end{subfigure}
\begin{subfigure}[t]{0.35\textwidth}
\includegraphics[width=\textwidth]{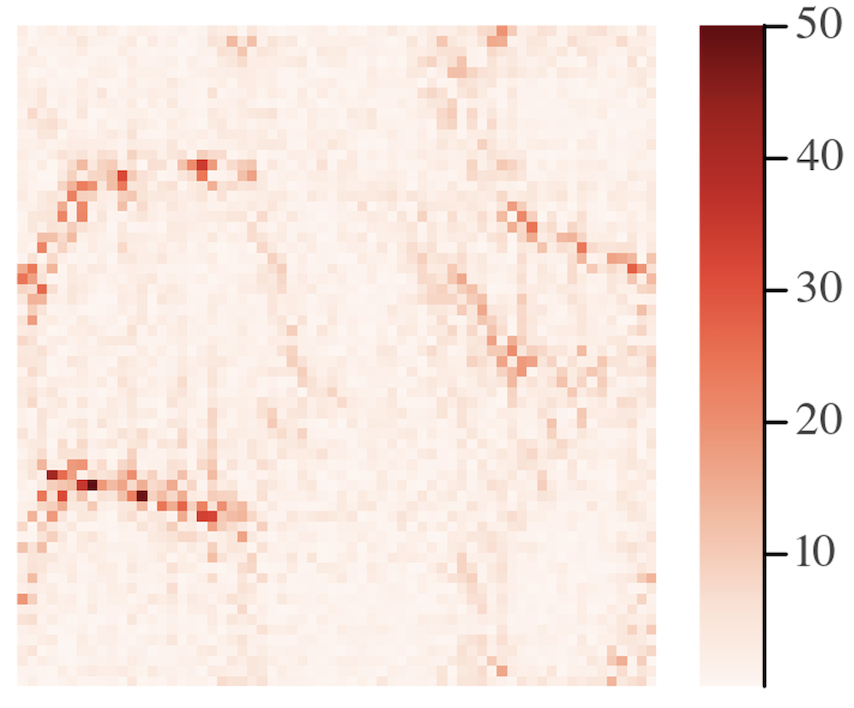}
\caption{\hbox{PiFNO, \(c=10^{-2}\)}}
\end{subfigure}
\hspace{6mm}
\begin{subfigure}[t]{0.35\textwidth}
\includegraphics[width=\textwidth]{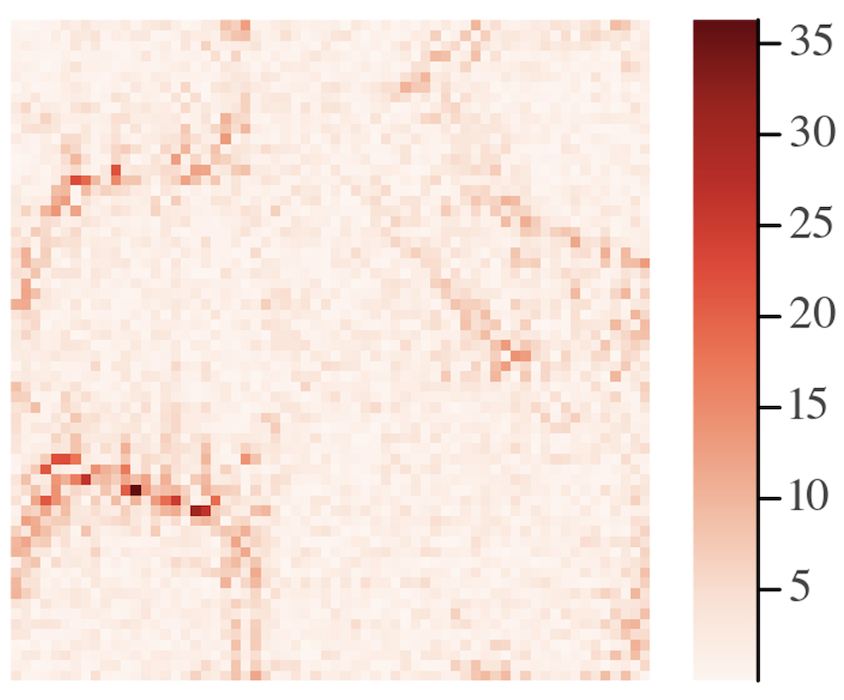}
\caption{\hbox{PiFNO, \(c=10^{-1}\)}}
\end{subfigure}
\caption{Trained FNO results for 
\(\mathop{\mathrm{div}}\tilde{\mathbf{p}}_{1}^{\mathrm{out}}\) [MPa/\(l\)]. 
Note the differences in scale. 
}
\label{fig:P22_P23_PDEError_comparison}
\end{figure}

\begin{figure}[H]
\centering
\begin{subfigure}[t]{0.35\textwidth}
\includegraphics[width=\textwidth]{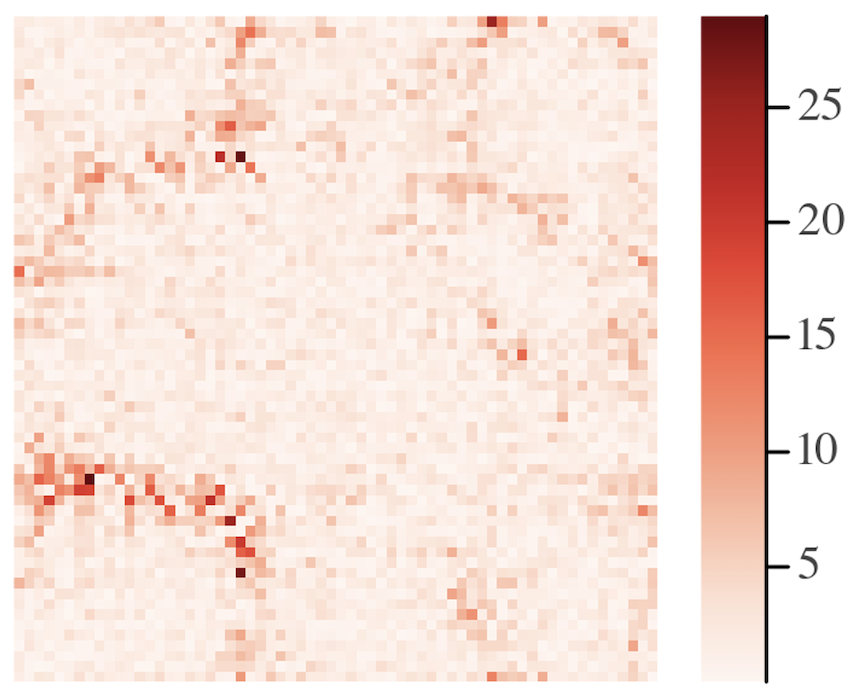}
\caption{\hbox{PgFNO}}
\vspace{3mm}
\end{subfigure}
\hspace{6mm}
\begin{subfigure}[t]{0.4\textwidth}
\includegraphics[width=\textwidth]{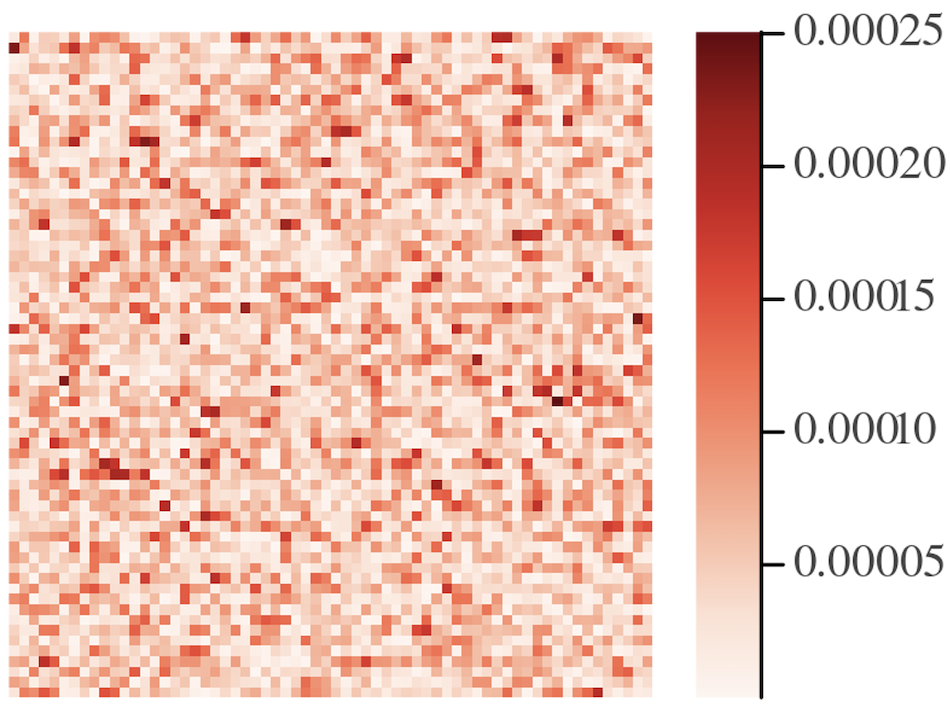}
\caption{\hbox{PeFNO}}
\vspace{3mm}
\end{subfigure}
\begin{subfigure}[t]{0.35\textwidth}
\includegraphics[width=\textwidth]{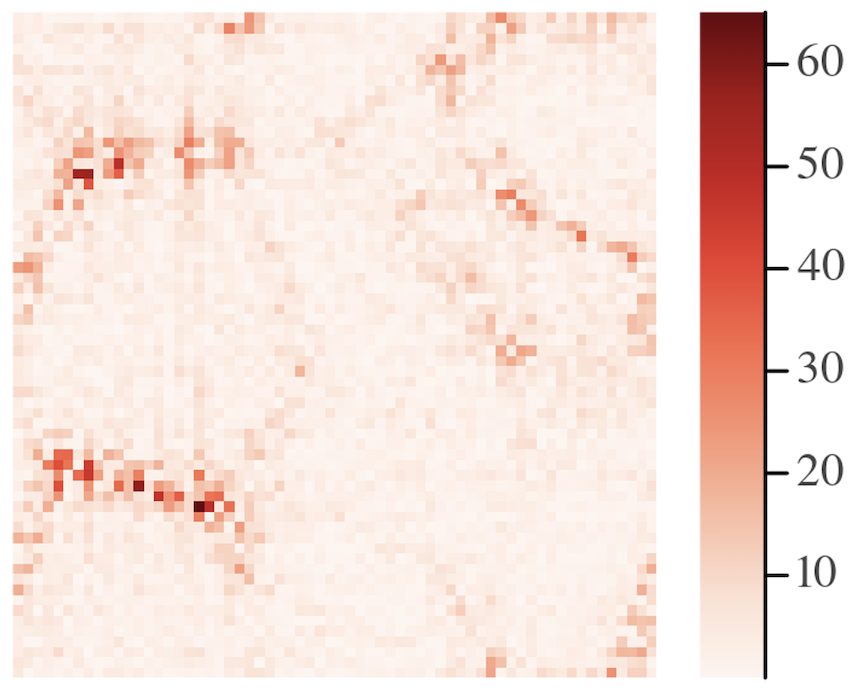}
\caption{\hbox{PiFNO, \(c=10^{-2}\)}}
\end{subfigure}
\hspace{6mm}
\begin{subfigure}[t]{0.35\textwidth}
\includegraphics[width=\textwidth]{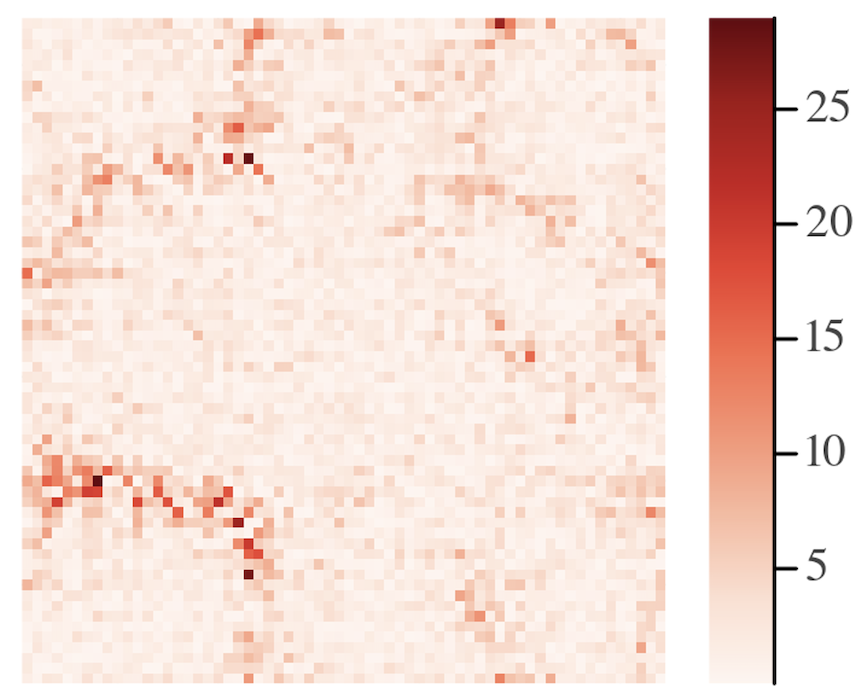}
\caption{\hbox{PiFNO, \(c=10^{-1}\)}}
\end{subfigure}
\caption{Trained FNO results for 
\(\mathop{\mathrm{div}}\tilde{\mathbf{p}}_{2}^{\mathrm{out}}\) [MPa/\(l\)]. 
Note the differences in scale. 
}
\label{fig:P32_P33_PDEError_comparison}
\end{figure}

To display these results, we work with the form 
\begin{equation}
\mathop{\mathrm{div}}\mathbf{p}_{r}
=\mathop{\mathrm{div}}\tilde{\mathbf{p}}_{r}
=\tilde{P}_{r1,1}+\tilde{P}_{r2,2}+\tilde{P}_{r3,3}
=0
\,,\ \ 
r=1,2,3
\,,
\end{equation}
of quasi-static mechanical equilibrium \Cref{equ:EquMecStaQua},  
where \(\mathbf{p}_{r}:=(P_{r1},P_{r2},P_{r3})\) is the \(r^{\mathrm{th}}\) 
row of \(\mathbf{P}\) from \Cref{equ:KirPioFirComCar}. Since for 
the current example of plane deformation in \((x_{1},x_{2})\), 
\(\mathop{\mathrm{div}}\tilde{\mathbf{p}}_{3}=0\) is identically zero, 
attention is focused on 
\(\mathop{\mathrm{div}}\tilde{\mathbf{p}}_{1}^{\mathrm{out}}\) 
and \(\mathop{\mathrm{div}}\tilde{\mathbf{p}}_{2}^{\mathrm{out}}\) 
here. Corresponding results are given in 
in Figures \ref{fig:P22_P23_PDEError_comparison} 
and \ref{fig:P32_P33_PDEError_comparison}. 
As expected, the values of 
\(\mathop{\mathrm{div}}\tilde{\mathbf{p}}_{1}^{\mathrm{out}}\) 
and  \(\mathop{\mathrm{div}}\tilde{\mathbf{p}}_{2}^{\mathrm{out}}\) 
calculated by the trained PeFNO are significantly smaller than those calculated 
by the other FNOs. This is also the case for (the magnitude of) 
\(\mathop{\mathrm{div}}\mathbf{P}^{\mathrm{out}}\) since 
\(|\mathop{\mathrm{div}}\mathbf{P}^{\mathrm{out}}|
=\sum_{r=1}^{3}
|\mathop{\mathrm{div}}\tilde{\mathbf{p}}_{r}^{\mathrm{out}}|\). 
Note also that the magnitude of the errors in 
\(\mathop{\mathrm{div}}\tilde{\mathbf{p}}_{1}^{\mathrm{out}}\) 
and \(\mathop{\mathrm{div}}\tilde{\mathbf{p}}_{2}^{\mathrm{out}}\) 
calculated by the trained PgFNO and PiFNO are similar and largest 
at the grain boundaries with the largest contrast in material properties 
(cf.~Figures \ref{fig:ProMatDat} and \ref{fig:WeiFit}). One possibility 
to reduce such error would be to increase the field resolution 
(i.e., \(n_{\mathrm{dis}}\)), especially near grain boundaries and more 
generally in 
regions of large contrasts in material properties. Another could be to  
train the FNOs with data on the stress field and its gradient field, at the 
cost of course of increased effort. 

\section{Summary and outlook} 
\label{sec:OutSum}

A physics-encoded Fourier neural operator (PeFNO) has been developed 
in the current work for the surrogate modeling of quasi-static equilibrium 
stress field in solids. For the corresponding constraint of divergence-free stress, 
a novel encoding approach based on a stress potential is proposed. 
As shown in the current work, inclusion ("encoding") of this 
constraint in the operator architecture rather than in the loss function 
as done in the physics-informed case 
yields more accurate and robust FNO output for the equillibrium stress field. 
This is also related to the fact that only the training of the PiFNO is 
constrained by mechanical equilibrium; in contrast, it constrains both 
the training and output of the PeFNO. 

The neural operators considered in the current work can be improved 
and refined in a number of ways. One possibility here is inclusion of 
the deformation gradient \(\mathbf{F}\) in the training and operator 
input data. Another concerns the architecture of the PeFNO for 
divergence-free stress based on the stress potential \(\mathbf{A}\). 
Since \(\tilde{\mathbf{P}}=\mathop{\mathrm{curl}}\tilde{\mathbf{A}}\) 
is not invertible, calculating \(\tilde{\mathbf{A}}\) as the 
"anti-curl" of \(\tilde{\mathbf{P}}\) is non-unique and a further 
source of error in operator training. 
To avoid this, further constraints on \(\tilde{\mathbf{A}}\) are necessary. 
The most common of these is the Coulomb gauge condition 
\(\mathop{\mathrm{div}}\tilde{\mathbf{A}}=\bm{0}\). 
In this case, \(\tilde{\mathbf{P}}=\mathop{\mathrm{curl}}\tilde{\mathbf{A}}\) 
is invertible, and \(\bar{\mathbf{P}}=\bar{\mathbf{A}}\) need not be learned, 
but rather holds identically. A corresponding improved PeFNO 
architecture is then obtained by adding an additional input layer based on 
the Coulomb gauge condition. These and other possible 
further development of the PeFNOs 
for surrogate computational modeling in solid mechanical represent 
work in progress to be reported on in the future. 

\bibliographystyle{elsarticle-harv}

\bibliography{References}

%\end{document}

\clearpage
%\nolinenumbers
\thispagestyle{empty}

% numbering 
\renewcommand{\thepage}{S.\arabic{page}}
\setcounter{page}{1}
\renewcommand{\thesection}{SI.\arabic{section}}
\setcounter{section}{0}
\renewcommand{\thesubsection}{SI.\arabic{section}.\arabic{subsection}}
\setcounter{subsection}{0}
\renewcommand{\theequation}{SI.\arabic{equation}}
\setcounter{equation}{0}

\begin{center}
\textrm{\large Supplementary Information}
\\[5mm]
\textbf{\Large A physics-encoded Fourier neural operator approach 
for surrogate modeling of divergence-free stress fields in solids} 
\\[5mm]
\textrm{Mohammad S. Khorrami$^{1,*}$, 
Pawan Goyal$^{2}$,
Jaber R. Mianroodi$^{1}$,\\
Bob Svendsen$^{1,3}$, Peter Benner$^{2}$, Dierk Raabe$^{1}$}
\\[3mm]
\textrm{\footnotesize 
${}^{1}$Microstructure Physics and Alloy Design,\\
Max-Planck Institute for Sustainable Materials, 
D\"usseldorf, Germany
\\[2mm]
${}^{2}$Computational Methods in Systems and Control Theory,\\ 
Max-Planck Institute for Dynamics of Complex Technical Systems, 
Magdeburg, Germany
\\[2mm]
${}^{3}$Material Mechanics, RWTH Aachen University, Aachen, Germany
\\[1mm]
${}^{*}$Corresponding Author: m.khorrami@mpie.de} 
\end{center}

\section{Introduction} 

The purpose of the Supplementary Information is to briefly summarize 
and document the mathematical background and details relevant to the 
current work, and in particular to the encoding of divergence-free stress 
fields in a FNO-based architecture. 

As discussed in the paper, encoding in the current approach 
is based in particular on a 
stress potential. The mathematical basis for such potentials is 
the Helmholtz decomposition of vector fields \cite[e.g.,][]{Bhatia2013}. 
This is briefly summarized in Section \ref{sec:DecHH} and compared 
with the related Hodge decomposition of 1-form fields employed 
by \cite{RichterPowell2022} to encode divergence-free vector fields in 
NN architectures. Central to the current work is a 
generalization of the Helmholtz decomposition for vector fields 
to second-order tensor fields also treated in Section \ref{sec:DecHH}. 

Application of the Helmholtz decomposition to stress measures 
and quasi-static mechanical equilibrium is documented in 
Section \ref{sec:PotStsFPK}. For solids, the former include the 
non-linear first Piola-Kirchhoff (PK) stress in the paper as well 
as the linear symmetric stress. 
Fourier series forms of the corresponding potential relations and 
other relations relevant to the development of the PeFNO are 
also given in Section \ref{sec:PotStsFPK}. 

The Supplementary Information ends in Section \ref{sec:ConCom} 
with a summary of the computational forms of the 
relations from Section \ref{sec:PotStsFPK} which have been 
implemented in the current PeFNO. Included here in particular 
is a brief technical discussion of the discrete Fourier transform (DFT) 
for real-valued fields (RDFT). This latter represents the 
basis of the real-valued fast Fourier transform 
\cite[RFFT:~e.g.,][]{Sorensen1987} employed in the FNO 
\cite[e.g.,][]{Li2021}. 

The mathematical treatment in the sequel employs the following notation.  
Three-dimensional Euclidean points or vectors are symbolized by lower case 
bold italic or greek letters \(\bm{a},\ldots,\bm{\zeta}\), 
and second-order Euclidean tensors by upper case bold italic or greek 
letters \(\bm{A},\ldots,\bm{Z}\). In particular, let 
\(\bm{i}_{1},\bm{i}_{2},\bm{i}_{3}\) represent the Cartesian basis 
vectors, and \(\bm{I}\) the identity. The notation \(\bm{a}\cdot\bm{b}\) 
and \(\bm{A}\cdot\bm{B}\) denotes scalar products. 
The transpose \(\bm{A}^{\!\mathrm{T}}\) of \(\bm{A}\) is defined by 
\(\bm{c}\cdot\bm{A}^{\!\mathrm{T}}\bm{b}:=\bm{b}\cdot\bm{A}\bm{c}\). 
The tensor product \(\bm{a}\otimes\bm{b}\) is defined by 
\((\bm{a}\otimes\bm{b})\bm{c}:=(\bm{b}\cdot\bm{c})\,\bm{a}\), and 
\((\mathop{\mathrm{axt}}\bm{a})\,\bm{b}:=\bm{a}\times\bm{b}\) 
defines the axial tensor \(\mathop{\mathrm{axt}}\bm{a}\) of any vector 
\(\bm{a}\). Additional concepts and notation are introduced as needed. 

\section{Helmholtz \& Hodge decompositions} 
\label{sec:DecHH}

For simplicity, attention is restricted here to fields on unbounded domains. 

\subsection{Helmholtz decomposition of vector fields} 

Let \(\nabla\) represent the Euclidean gradient operator, and 
\(\nabla\bm{u}\) the gradient of a differentiable vector field \(\bm{u}\). 
Likewise, let \(\mathop{\mathrm{div}}\bm{u}\) represent the 
divergence, and \(\mathop{\mathrm{curl}}\bm{u}\) the curl, 
of \(\bm{u}\) defined by\footnote{Unless restricted by parentheses, 
all operators apply to everything on their right.} 
\begin{equation}
\mathop{\mathrm{div}}\bm{u}
:=\bm{I}\cdot\nabla\bm{u}
\,,\quad
\bm{c}\cdot\mathop{\mathrm{curl}}\bm{u}
:=\mathop{\mathrm{div}}\bm{u}\times\bm{c}
=\mathop{\mathrm{div}}\,(\mathop{\mathrm{axt}}\bm{u})\,\bm{c}
\,,
\label{equ:FieVecCurDivCalEuc}
\end{equation} 
\cite[e.g.,][, Chapter 1]{Chadwick1999} for constant \(\bm{c}\). 
Given these, the Helmholtz decomposition of \(\bm{u}\) takes the form 
\begin{equation}
\bm{u}=\nabla\varsigma+\mathop{\mathrm{curl}}\bm{\varphi}
\label{equ:FieVecDecHel}
\end{equation} 
\cite[e.g.,][]{Bhatia2013} 
determined by scalar \(\varsigma\) and vector \(\bm{\varphi}\) potentials. 
Properties of this split and the potentials include
\vspace{-3mm}
\begin{itemize}
\itemsep-1pt
\item \(\varsigma\) and \(\bm{\varphi}\) are determined only up to constants, 
\item \(\bm{\varphi}+\nabla f\) is also a vector potential for any smooth 
scalar field \(f\), 
\item \(\bm{u}\) is divergence-free for \(\varsigma\) harmonic, i.e., 
\(\mathop{\mathrm{div}}\nabla\varsigma=0\). 
\end{itemize} 
\vspace{-2mm}
In particular, the trivial scalar potential 
\begin{equation}
\varsigma=\mathrm{const.}
\label{equ:FieVecPotScaTri}
\end{equation}
determines the corresponding special case 
\begin{equation}
\bm{u}=\mathop{\mathrm{curl}}\bm{\varphi}
\label{equ:FieVecFreDiv}
\end{equation} 
of \eqref{equ:FieVecDecHel} relevant to the encoding of divergence-free 
vector fields in NN and NO architectures. 

\subsection{Hodge decomposition of 1-form fields} 

To encode divergence-free vector fields into NN architectures, 
\cite{RichterPowell2022} work with 
the Hodge decomposition of 1-form fields \cite[e.g.,][]{Abr88} 
rather than with \eqref{equ:FieVecDecHel}. 
Although they consider the \(n\)-dimensional case, 
for comparison with the current approach, 
attention is restricted here for simplicity to the physically relevant case \(n=3\). 

Let \(g\) represent the standard Euclidean metric with 
\(g(\bm{a},\bm{b})=\bm{a}\cdot\bm{b}\), 
and \(\omega\) the standard 
Euclidean triple product with \(\omega(\bm{a},\bm{b},\bm{c})
=g(\bm{a}\times\bm{b},\bm{c})\). 
The interior product operator \(\imath_{\bm{a}}\) \cite[e.g.,][, \S 5.1]{Abr88} 
maps \(g\) to the 1-covector \(\imath_{\bm{a}}g\) defined by 
\((\imath_{\bm{a}}g)(\bm{b}):=g(\bm{a},\bm{b})\), and \(\omega\) 
to the 2-covector \(\imath_{\bm{a}\,}\omega\) defined by 
\((\imath_{\bm{a}\,}\omega)(\bm{b},\bm{c}):=\omega(\bm{a},\bm{b},\bm{c})\). 
The Hodge star operator \(\star\) \cite[e.g.,][, Definition 6.2.12]{Abr88} 
maps \(k\)-covectors (\(k=0,\ldots,3\)) to \(3-k\) covectors and 
\(\star\star=(-1)^{k(3-k)}\). In particular, note that \(\star 1=\omega\), 
\(\star\omega=1\), 
\(\star\imath_{\bm{c}}g=\imath_{\bm{c}\,}\omega\) and 
\(\star\imath_{\bm{c}\,}\omega=\imath_{\bm{c}}g\). 
In terms of these operations, the Hodge decomposition of the 1-form 
field \(\imath_{\bm{u}}g\) dual to \(\bm{u}\) takes the form 
\begin{equation}
\imath_{\bm{u}}g=d\varsigma+\star d\imath_{\bm{\varphi}}g
=\imath_{\nabla\varsigma}g+\imath_{\mathop{\mathrm{curl}}\bm{\varphi}}g
\,,
\label{equ:FieForDecHog}
\end{equation} 
where \(\varsigma\) and \(\bm{\varphi}\) are the same potentials as 
in \eqref{equ:FieVecDecHel}, and \(d\) is the exterior derivative operator 
\cite[e.g.,][, Chapter 6]{Abr88}. To be precise, 
\cite{RichterPowell2022} work with the axial tensor 
\(\mathop{\mathrm{axt}}\bm{\varphi}\) of \(\bm{\varphi}\) 
(their \(\bm{A}\)) as well as the alternative vector potential \(\bm{b}\) with 
\(\star d\!\star\!\imath_{\bm{b}\,}\omega=\imath_{\bm{\varphi}}g\). 
Choice of the trivial scalar potential \eqref{equ:FieVecPotScaTri} reduces 
\eqref{equ:FieForDecHog} to 
\begin{equation}
\imath_{\bm{u}}g=\star d\imath_{\bm{\varphi}}g
\label{equ:FieForFreDiv}
\end{equation} 
which is identically divergence-free since then 
\(
\mathop{\mathrm{div}}\bm{u}
=\star d\imath_{\bm{u}\,}\omega
=\star dd\imath_{\bm{\varphi}}g
=0\). 

For vector fields \(\bm{u}\) and their 1-form duals 
\(\imath_{\bm{u}}g\), the Helmholtz \eqref{equ:FieVecDecHel} 
and Hodge \eqref{equ:FieForDecHog} decompositions are clearly 
equivalent. In contrast to the Helmholtz case, however, the Hodge 
decomposition is not readily generalizable 
to higher- and in particular second-order tensor fields like the stress. 

\subsection{Helmholtz decomposition for tensor fields}

This is obtained from the vector form \eqref{equ:FieVecDecHel} as follows. 
Any differentiable second-order tensor field \(\bm{S}\) 
and constant vector \(\bm{c}\) induce a corresponding vector field 
\(\bm{v}=\bm{S}^{\mathrm{T}}\!\bm{c}\) whose divergence 
and curl are defined by \eqref{equ:FieVecCurDivCalEuc}. Using these, 
one can define the corresponding operators 
\begin{equation}
\bm{c}\cdot\mathop{\mathrm{div}}\bm{S}
:=\mathop{\mathrm{div}}\bm{S}^{\mathrm{T}}\!\bm{c}
\,,\quad
(\mathop{\mathrm{curl}}\bm{S})^{\mathrm{T}}\bm{c}
:=\mathop{\mathrm{curl}}\bm{S}^{\mathrm{T}}\!\bm{c}
\,,
\label{equ:FieTenCurDivCalEuc}
\end{equation} 
on \(\bm{S}\). 
Analogously, substitution of 
\(\bm{u}=\bm{S}^{\mathrm{T}}\bm{c}\), 
\(\varsigma=\bm{\phi}\cdot\bm{c}\), and 
\(\bm{\varphi}=\bm{\Phi}^{\mathrm{T}}\bm{c}\) 
into the vector Helmholtz decomposition \eqref{equ:FieVecDecHel} 
yields its generalization to second-order tensor fields 
\begin{equation}
\bm{S}=\nabla\bm{\phi}+\mathop{\mathrm{curl}}\bm{\Phi}
\label{equ:FieTenHelSto}
\end{equation} 
via \eqref{equ:FieTenCurDivCalEuc}${}_{2}$ and the identity 
\(\nabla(\bm{\phi}\cdot\bm{c})=(\nabla\bm{\phi})^{\mathrm{T}}\bm{c}\), 
again for constant \(\bm{c}\). The PeFNO in the paper 
for divergence-free stress employs the special case 
\begin{equation}
\bm{\phi}=\mathrm{const.}
\quad\Longrightarrow\quad
\bm{S}=\mathop{\mathrm{curl}}\bm{\Phi}
\label{equ:FieTenGenFreDiv}
\end{equation} 
of \eqref{equ:FieTenHelSto} analogous to \eqref{equ:FieVecFreDiv}. 

\subsection{Mean-fluctuation split of fields on a region} 

The PeFNO in the paper is developed for stress fields on a 
(in general three-dimensional) spatial region \(U\). On such a region, 
note that any integrable field \(f\) can be split into a sum 
\begin{equation}
\textstyle
f(\bm{x})
=\bar{f}+\tilde{f}(\bm{x})
\,,\quad
\bar{f}
:=\frac{1}{v(U)}
\int_{U}f(\bm{x})\ dv(\bm{x})
\,,\ \ 
\tilde{f}(\bm{x})
:=f(\bm{x})-\bar{f}
\,,
\label{equ:SplFluMea}
\end{equation} 
of mean \(\bar{f}\) and fluctuation \(\tilde{f}\) parts, where 
\(v(U)=\int_{U}dv(\bm{x})\) is the volume of \(U\). 
In particular, then, \(\bm{S}=\skew5\bar{\bm{S}}+\skew5\tilde{\bm{S}}\) 
and \(\bm{\Phi}=\skew4\bar{\bm{\Phi}}+\skew4\tilde{\bm{\Phi}}\) 
for the tensor fields in \eqref{equ:FieTenGenFreDiv}. Since their mean parts 
are constant, note that 
\(\mathop{\mathrm{div}}\bm{S}
=\mathop{\mathrm{div}}\skew5\tilde{\bm{S}}\) and 
\(\mathop{\mathrm{curl}}\bm{\Phi}
=\mathop{\mathrm{curl}}\skew4\tilde{\bm{\Phi}}\), 
reducing \eqref{equ:FieTenGenFreDiv} to 
\begin{equation}
\skew5\tilde{\bm{S}}=\mathop{\mathrm{curl}}\skew4\tilde{\bm{\Phi}}
\label{equ:FieTenFreDiv}
\end{equation} 
in the context of \eqref{equ:SplFluMea}.

\section{Application to divergence-free stress} 
\label{sec:PotStsFPK}

\subsection{Stress potential field relations} 

The potential relation \eqref{equ:FieTenFreDiv} is directly applicable 
to non-linear quasi-static mechanical equilibrium in solids based on the first 
PK stress \(\bm{P}\), i.e., 
\begin{equation}
\skew5\tilde{\bm{P}}=\mathop{\mathrm{curl}}\skew5\tilde{\bm{A}}
\,,
\label{equ:PotStsFPK}
\end{equation}
where \(\bm{A}\) is the potential for \(\bm{P}\). In contrast to \(\bm{P}\), 
the linear stress tensor \(\bm{T}\) is symmetric 
(i.e., \(\bm{T}^{\mathrm{T}}=\bm{T}\)), 
and \eqref{equ:FieTenFreDiv} is not directly applicable. 
To preserve the symmetry of \(\bm{T}\), \eqref{equ:FieTenFreDiv} 
must be modified by the choice \(\skew4\tilde{\bm{\Phi}}
=(\mathop{\mathrm{curl}}\skew4\tilde{\bm{B}})^{\mathrm{T}}\) 
with \(\bm{B}\) symmetric. The resulting potential relation 
\begin{equation}
\skew3\tilde{\bm{T}}
=\mathop{\mathrm{inc}}\skew4\tilde{\bm{B}}
:=\mathop{\mathrm{curl}}
\,(\mathop{\mathrm{curl}}\skew4\tilde{\bm{B}})^{\mathrm{T}}
\label{equ:PotStsBel}
\end{equation}
for \(\bm{T}\) is then identically symmetric (as evident for example from 
\eqref{equ:StsKirPioFirPotSerForCur}${}_{3}$ below) and divergence-free. In 
linear elasticity, the operator \(\mathop{\mathrm{inc}}\) is known 
as the incompatibility operator \cite[e.g.,][]{Teo82}, and 
\(\bm{B}\) is known as the Beltrami stress potential or "function"  
\cite[e.g.,][, \S13.6]{Sadd2009}. 

\subsection{Fourier series relations} 

Assume next that all fields are periodic on \(U\). The Fourier series of 
any field \(f\) integrable and periodic on \(U\) is given by 
\begin{equation}
\textstyle
f(\bm{x})
=\sum_{\bm{k}\in U^{\ast}}
e_{}^{\imath{}\bm{k}\cdot\bm{x}}
\hat{f}(\bm{k})
\,,\quad
\hat{f}(\bm{k})
=\frac{1}{v(U)}
\int_{U}
e_{}^{-\imath{}\bm{k}\cdot\bm{x}}
\,f(\bm{x})
\ dv(\bm{x})
\,,
\label{equ:FieScaSerFor}
\end{equation} 
where \(\imath=\sqrt{-1}\) and \(U^{\ast}\) is the wavevector 
space of \(U\) (examples given below). In particular, then 
\begin{equation}
\begin{array}{rclcrcl}
\bm{P}(\bm{x})
&=&
\sum_{\bm{k}\in U^{\ast}}
e_{}^{\imath\bm{k}\cdot\bm{x}}
\skew3\hat{\bm{P}}(\bm{k})
\,,&&
\bm{A}(\bm{x})
&=&
\sum_{\bm{k}\in U^{\ast}}
e_{}^{\imath\bm{k}\cdot\bm{x}}
\skew5\hat{\bm{A}}(\bm{k})
\,,\\
\bm{T}(\bm{x})
&=&
\sum_{\bm{k}\in U^{\ast}}
e_{}^{\imath\bm{k}\cdot\bm{x}}
\skew2\hat{\bm{T}}(\bm{k})
\,,&&
\bm{B}(\bm{x})
&=&
\sum_{\bm{k}\in U^{\ast}}
e_{}^{\imath\bm{k}\cdot\bm{x}}
\skew5\hat{\bm{B}}(\bm{k})
\,.
\end{array}
\label{equ:PotStsSerFor}
\end{equation} 
The mean-fluctuation split 
\eqref{equ:SplFluMea} takes the Fourier series form
\begin{equation}
\textstyle
f(\bm{x})
=\bar{f}+\tilde{f}(\bm{x})
\,,\quad
\bar{f}
=\hat{f}(\bm{0})
\,,\ \ 
\tilde{f}(\bm{x})
=\sum_{\bm{k}\neq\bm{0}}
e_{}^{\imath{}\bm{k}\cdot\bm{x}}
\hat{f}(\bm{k})
\,,
\label{equ:SplFluMeaSerFou}
\end{equation}
with \(\sum_{\bm{k}\neq\bm{0}}
:=\sum_{\bm{k}\in U^{\ast}\setminus\lbrace\bm{0}\rbrace}\). 
This implies in particular (recall that \(\bm{B}\) is symmetric) 
\begin{equation}
\begin{array}{rclcl}
\mathop{\mathrm{div}}\bm{P}(\bm{x})
&=&
\mathop{\mathrm{div}}\skew3\tilde{\bm{P}}(\bm{x})
&=&
\sum_{\bm{k}\neq\bm{0}}
e_{}^{\imath{}\bm{k}\cdot\bm{x}}
\skew3\hat{\bm{P}}(\bm{k})
\,\imath\bm{k}
\,,\\
\mathop{\mathrm{div}}\bm{T}(\bm{x})
&=&
\mathop{\mathrm{div}}\skew2\tilde{\bm{T}}(\bm{x})
&=&
\sum_{\bm{k}\neq\bm{0}}
e_{}^{\imath{}\bm{k}\cdot\bm{x}}
\skew2\hat{\bm{T}}(\bm{k})
\,\imath\bm{k}
\,,\\
\mathop{\mathrm{curl}}\bm{A}(\bm{x})
&=&
\mathop{\mathrm{curl}}\skew5\tilde{\bm{A}}(\bm{x})
&=&
\sum_{\bm{k}\neq\bm{0}}
e_{}^{\imath{}\bm{k}\cdot\bm{x}}
\skew5\hat{\bm{A}}(\bm{k})
\,(\mathop{\mathrm{axt}}\imath\bm{k})^{\mathrm{T}}
\,,\\
\mathop{\mathrm{inc}}\bm{B}(\bm{x})
&=&
\mathop{\mathrm{inc}}\skew4\tilde{\bm{B}}(\bm{x})
&=&
\sum_{\bm{k}\neq\bm{0}}
e_{}^{\imath{}\bm{k}\cdot\bm{x}}
(\mathop{\mathrm{axt}}\imath\bm{k})
\,\skew4\hat{\bm{B}}(\bm{k})
\,(\mathop{\mathrm{axt}}\imath\bm{k})^{\mathrm{T}}
\,.
\end{array}
\label{equ:StsKirPioFirPotSerForCur}
\end{equation} 
In the context of the potential relations \eqref{equ:PotStsFPK} 
and \eqref{equ:PotStsBel}, then, 
\begin{equation}
\begin{array}{rcl}
\skew3\hat{\bm{P}}(\bm{k})
&=&
\left\lbrace
\begin{array}{lcl}
\skew5\hat{\bm{A}}(\bm{k})
&&
\bm{k}=\bm{0}
\\
\skew5\hat{\bm{A}}(\bm{k})
\,(\mathop{\mathrm{axt}}\imath\bm{k})^{\mathrm{T}}
&&
\bm{k}\neq\bm{0}
\end{array}
\right.
\,,
\\[5mm]
\skew3\hat{\bm{T}}(\bm{k})
&=&
\left\lbrace
\begin{array}{lcl}
\skew4\hat{\bm{B}}(\bm{k})
&&
\bm{k}=\bm{0}
\\
(\mathop{\mathrm{axt}}\imath\bm{k})
\,\skew4\hat{\bm{B}}(\bm{k})
\,(\mathop{\mathrm{axt}}\imath\bm{k})^{\mathrm{T}}
&&
\bm{k}\neq\bm{0}
\end{array}
\right.
\,,
\end{array}
\label{equ:PotStsCofFor}
\end{equation} 
respectively. These also follow from the fact that, since the potential 
relations involve only the fluctuation parts of the corresponding fields, 
\(\skew3\hat{\bm{P}}(\bm{0})=\skew5\hat{\bm{A}}(\bm{0})\) 
and \(\skew3\hat{\bm{T}}(\bm{0})=\skew4\hat{\bm{B}}(\bm{0})\) 
may be assumed without loss of physical generality. 

\section{Computational details}
\label{sec:ConCom}

\subsection{Cartesian component relations}

For brevity and simplicity, the treatment in the paper is limited to the 
non-linear case \eqref{equ:PotStsFPK} and Cartesian component relations. 
In particular, 
\begin{equation}
\textstyle
\bm{P}=\sum_{i=1}^{3}\sum_{j=1}^{3}
P_{\!ij\,}\bm{i}_{i}\otimes\bm{i}_{j}
\,,\quad
\bm{A}=\sum_{i=1}^{3}\sum_{j=1}^{3}
A_{ij\,}\bm{i}_{i}\otimes\bm{i}_{j}
\,.
\end{equation}
These result in the Cartesian component matrix forms 
\begin{equation}
\textstyle
\mathbf{P}(\mathbf{x})
=\sum_{\mathbf{k}\in\mathrm{U}^{\ast}}
e^{\imath\mathbf{k}\cdot\mathbf{x}}
\,\hat{\mathbf{P}}(\mathbf{k})
\,,\ \ 
\mathbf{A}(\mathbf{x})
=\sum_{\mathbf{k}\in\mathrm{U}^{\ast}}
e^{\imath\mathbf{k}\cdot\mathbf{x}}
\hat{\mathbf{A}}(\mathbf{k})
\,,
\label{equ:PotStsRepFouMatCar}
\end{equation} 
of \eqref{equ:PotStsSerFor}${}_{1,2}$, respectively, 
with \(\mathbf{k}:=(k_{1},k_{2},k_{3})\) and 
\(\mathbf{x}:=(x_{1},x_{2},x_{3})\). In particular, 
the output layer in the architecture (light red in \Cref{fig:ArcANcP}(a)) 
of the PgFNO and PiFNO is based on \eqref{equ:PotStsRepFouMatCar}${}_{1}$. 
Evaluation of the PiFNO loss function \Cref{equ:FunLosInfPhy} is based on 
the Cartesian component array form 
\begin{equation}
\textstyle
\mathop{\mathrm{div}}\mathbf{P}(\mathbf{x})
=\sum_{\mathbf{k}\neq\bm{0}}
e^{\imath\mathbf{k}\cdot\mathbf{x}}
\,\skew4\hat{\mathbf{d}}(\mathbf{k})
\label{equ:StsDivRepFouAryCar}
\end{equation} 
of \eqref{equ:StsKirPioFirPotSerForCur}${}_{1}$ with 
\begin{equation}
\skew4\hat{\mathbf{d}}(\mathbf{k})
:=\hat{\mathbf{P}}(\mathbf{k})
\,\imath\mathbf{k}
=\imath
\left\lbrack
\begin{array}{c}
\hat{P}_{11}(\mathbf{k})\,k_{1}
+\hat{P}_{12}(\mathbf{k})\,k_{2}
+\hat{P}_{13}(\mathbf{k})\,k_{3}
\\
\hat{P}_{21}(\mathbf{k})\,k_{1}
+\hat{P}_{22}(\mathbf{k})\,k_{2}
+\hat{P}_{23}(\mathbf{k})\,k_{3}
\\
\hat{P}_{31}(\mathbf{k})\,k_{1}
+\hat{P}_{32}(\mathbf{k})\,k_{2}
+\hat{P}_{33}(\mathbf{k})\,k_{3}
\end{array}
\right\rbrack
.
\end{equation} 
The output transformation 
\(\mathbf{A}^{\!\mathrm{out}}\to\mathbf{P}^{\mathrm{out}}\) 
in the architecture (light red in \Cref{fig:ArcANcP}(b)) for 
the PeFNO is determined by the Cartesian component matrix form 
\begin{equation}
\textstyle
\hat{\mathbf{P}}(\mathbf{k})
=\left\lbrace
\begin{array}{lcl}
\hat{\mathbf{A}}(\mathbf{k})
&&
\mathbf{k}=\bm{0}
\\
\hat{\mathbf{A}}(\mathbf{k})
\,(\mathop{\mathrm{axt}}\imath\mathbf{k})^{\mathrm{T}}
&&
\mathbf{k}\neq\bm{0}
\end{array}
\right.
,
\label{equ:PotStsRepFouCof}
\end{equation} 
of \eqref{equ:PotStsCofFor}${}_{1}$ with 
\begin{equation}
\begin{array}{l}
\hat{\mathbf{A}}(\mathbf{k})
\,(\mathop{\mathrm{axt}}\imath\mathbf{k})^{\mathrm{T}}
\\
\quad=\ \ \imath
\left\lbrack
\begin{array}{ccc}
k_{2}\hat{A}_{13}(\mathbf{k})-k_{3}\hat{A}_{12}(\mathbf{k})&
k_{3}\hat{A}_{11}(\mathbf{k})-k_{1}\hat{A}_{13}(\mathbf{k})&
k_{1}\hat{A}_{12}(\mathbf{k})-k_{2}\hat{A}_{11}(\mathbf{k})
\\
k_{2}\hat{A}_{23}(\mathbf{k})-k_{3}\hat{A}_{22}(\mathbf{k})&
k_{3}\hat{A}_{21}(\mathbf{k})-k_{1}\hat{A}_{23}(\mathbf{k})&
k_{1}\hat{A}_{22}(\mathbf{k})-k_{2}\hat{A}_{21}(\mathbf{k})
\\
k_{2}\hat{A}_{33}(\mathbf{k})-k_{3}\hat{A}_{32}(\mathbf{k})&
k_{3}\hat{A}_{31}(\mathbf{k})-k_{1}\hat{A}_{33}(\mathbf{k})&
k_{1}\hat{A}_{32}(\mathbf{k})-k_{2}\hat{A}_{31}(\mathbf{k})
\end{array}
\right\rbrack
\label{equ:PotStsRepFouCofMat}
\end{array}
\end{equation} 
in the context of \eqref{equ:PotStsRepFouMatCar}. On this basis, 
note for example that 
\begin{equation}
\textstyle
\mathbf{P}_{\!ab}^{\mathrm{out}}
=\sum_{\mathbf{k}\in\mathrm{U}^{\ast}}
e^{\imath\mathbf{k}\cdot\mathbf{x}_{b}}
\,\hat{\mathbf{P}}_{\!a}^{\mathrm{out}}(\mathbf{k})
\,,\ \ 
(\mathop{\mathrm{div}}\mathbf{P}_{\!a}^{\mathrm{out}})_{b}
=\sum_{\mathbf{k}\neq\bm{0}}
e^{\imath\mathbf{k}\cdot\mathbf{x}_{b}}
\,\skew4\hat{\mathbf{d}}_{a}^{\mathrm{out}}(\mathbf{k})
\,,
\end{equation} 
hold for \(\mathbf{P}_{\!ab}^{\mathrm{out}}\) and 
\((\mathop{\mathrm{div}}\mathbf{P}_{\!a}^{\mathrm{out}})_{b}\), 
respectively, in \Cref{equ:FunLosEncGuiPhy} and \Cref{equ:FunLosInfPhy}. 

Implementation and numerical evaluation of these relations 
in the corresponding FNOs is based of course on their truncation and discretization. 

\subsection{Discrete Fourier transform for real-valued fields}

The multi-dimensional discrete Fourier transform (DFT) and its counterpart 
for real-valued fields (RDFT) represent tensor-product-based direct 
generalizations of the one-dimensional (1D) case. For this case, let 
\(U\) in \eqref{equ:FieScaSerFor} represent an interval of length \(l\). Then 
\begin{equation}
\begin{array}{rcl}
U
&=&
\lbrace
\bm{x}=x_{1}\bm{i}_{1}\ \vert\ x_{1}\in\lbrack 0,l\rbrack
\rbrace
\,,\\ 
U^{\ast}
&=&
\lbrace
\bm{k}=k_{1}\bm{i}_{1}
\ \vert\ k_{1}=2\pi\kappa_{1}/l,\,\kappa_{1}\ \hbox{integer}
\rbrace
\,,
\end{array}
\end{equation}
and \eqref{equ:FieScaSerFor} reduce to  
\begin{equation}
\textstyle
f(\xi_{1})
=\sum_{\kappa_{1}=-\infty}^{\infty}
e^{2\pi\imath\kappa_{1}\xi_{1}}
\hat{f}_{\kappa_{1}}
\,,\quad
\hat{f}_{\kappa_{1}}
=\int_{0}^{1}
e^{-2\pi\imath\kappa_{1}\xi_{1}}
f(\xi_{1})
\ d\xi_{1}
\,,
\label{equ:RepFieSerFor1D}
\end{equation} 
with \(\xi_{1}:=x_{1}/l\in\lbrack 0,1\rbrack\). 
Uniform discretization
\begin{equation}
\textstyle
\lbrack 0,1\rbrack
=\bigcup_{i_{1}=1}^{n_{1}}\lbrack \xi_{i_{1}},\xi_{i_{1}+1}\rbrack
\,,\quad
\xi_{i_{1}}:=(i_{1}-1)/n_{1}
\,,
\label{equ:DisUni1D}
\end{equation}
of \(\lbrack 0,1\rbrack\),
trapezoidal integration of \eqref{equ:RepFieSerFor1D}${}_{1}$, 
order \(m_{1}\) truncation of \eqref{equ:RepFieSerFor1D}${}_{2}$, 
and cardinality then determine the discretized form / approximation 
\begin{equation}
\begin{array}{rcl}
\check{f}_{\kappa_{1}}
&=&
\frac{1}{n_{1}}
\sum_{i_{1}=1}^{n_{1}}
\varphi_{\kappa_{1}i_{1}}^{-}
f_{i_{1}}
\,,\\[2mm]
f_{i_{1}}
&=&
\left\lbrace
\begin{array}{lcl}
\sum_{\kappa_{1}=-m_{1}}^{m_{1}}
\varphi_{\kappa_{1}i_{1}}^{+}
\check{f}_{\kappa_{1}}
&&
n_{1}\ \hbox{odd},\,m_{1}=\frac{1}{2}(n_{1}-1)
\\[1mm]
\sum_{\kappa_{1}=-m_{1}}^{m_{1}-1}
\varphi_{\kappa_{1}i_{1}}^{+}
\check{f}_{\kappa_{1}}
&&
n_{1}\ \hbox{even},\,m_{1}=\frac{1}{2}n_{1}
\end{array}
\right.
\,,
\end{array}
\label{equ:FieComSerForDis1D}
\end{equation} 
of \eqref{equ:RepFieSerFor1D} with 
\(\varphi_{\kappa_{1}i_{1}}^{\pm}
:=e^{\pm2\pi\imath\kappa_{1}(i_{1}-1)/n_{1}}\), 
\(\hat{f}_{\kappa_{1}}\approx \check{f}_{\kappa_{1}}\), and 
\(f_{i_{1}}:=f(\xi_{i_{1}})\). 

If \(f\) is real-valued, \(\hat{f}_{\kappa_{1}}\) in \eqref{equ:RepFieSerFor1D} 
satisfies in addition the symmetry condition 
\begin{equation}
\textstyle
\hat{f}_{\kappa_{1}}^{\ast}
=\int_{0}^{1}
e^{2\pi\imath\kappa_{1}\xi_{1}}
f(\xi_{1})
\ d\xi_{1}
=\hat{f}_{-\kappa_{1}}
\,,\quad
\kappa_{1}\neq 0
\,,
\end{equation}
via the identity \((z_{1}z_{2})^{\ast}=z_{1}^{\ast}z_{2}^{\ast}\), 
\(f^{\ast}=f\), and 
\((e^{-2\pi\imath\kappa_{1}\xi_{1}})^{\ast}
=e^{2\pi\imath\kappa_{1}\xi_{1}}\),
where \(z^{\ast}:=x-\imath y\) is the complex conjugate of \(z=x+\imath y\). 
This results in the reduced form 
\begin{equation} 
\begin{array}{rcl}
\check{f}_{\kappa_{1}}
&=&
\frac{1}{n_{1}}
\sum_{i_{1}=1}^{n_{1}}
\varphi_{\kappa_{1}i_{1}}^{-}
f_{i_{1}}
\,,\\[2mm]
f_{i_{1}}
&=&
\check{f}_{0}
+\left\lbrace
\begin{array}{lcl}
2\mathop{\mathrm{rea}}
\sum_{\kappa_{1}=1}^{m_{1}}
\varphi_{\kappa_{1}i_{1}}^{+}
\check{f}_{\kappa_{1}}
&&
n_{1}\ \hbox{odd},\,m_{1}=\frac{1}{2}(n_{1}-1)
\\[1mm]
2\mathop{\mathrm{rea}}
\sum_{\kappa_{1}=1}^{m_{1}-1}
\varphi_{\kappa_{1}i_{1}}^{+}
\check{f}_{\kappa_{1}}
+(-1)^{i_{1}+1}\check{f}_{\frac{1}{2}n_{1}}
&&
n_{1}\ \hbox{even},\,m_{1}=\frac{1}{2}n_{1}
\end{array}
\right.
\,,
\end{array}
\label{equ:FieReaSerForDis1D}
\end{equation}
of \eqref{equ:FieComSerForDis1D} 
with \(2\mathop{\mathrm{rea}}z=z+z^{\ast}\), where 
\(x=\mathop{\mathrm{rea}}z\) is the real part of \(z=x+\imath y\). 
In particular, for \(n_{1}\) even, 
\eqref{equ:FieReaSerForDis1D} is based on modal periodicity 
\(\check{f}_{\kappa_{1}\pm n_{1}}=\check{f}_{\kappa_{1}}\).

Comparing \eqref{equ:FieComSerForDis1D} (\(f\) complex-valued) 
and \eqref{equ:FieReaSerForDis1D} (\(f\) real-valued) 
for \(n_{1}\) even, for example, one sees that the former depends on 
\(2m_{1}\) modes, and the latter on \(m_{1}+1\) modes. 
This reduction in the number of independent modes represents the 
computational advantage of the RDFT and its fast version 
\cite[RFFT:~e.g.,][]{Sorensen1987}. For the discretization 
\(n_{1}=n_{\mathrm{dis}}=64\) employed in the paper, for example, 
the corresponding RFFT is based on \(\frac{1}{2}n_{1}+1=33\) 
independent modes. 

Proof-of-concept computations in the paper are carried out for plane 
deformation in a 2D square region / unit cell \(U\) of side length \(l\), 
i.e., 
\begin{equation}
\begin{array}{rcl}
U
&=&
\lbrace
\bm{x}=x_{1}\bm{i}_{1}+x_{2}\bm{i}_{2}
\ \vert\ 
x_{d}\in\lbrack 0,l\rbrack,
\,\ d=1,2
\rbrace
\,,\\
U^{\ast}
&=&
\lbrace
\bm{k}=k_{1}\bm{i}_{1}+k_{2}\bm{i}_{2}
\ \vert\ 
k_{d}=2\pi\kappa_{d}/l,\,\kappa_{d}\ \hbox{integer},
\,\ d=1,2
\rbrace
\,.
\end{array}
\label{equ:RepCarCelUniCom}
\end{equation} 
Tensor-product-based generalization of \eqref{equ:FieReaSerForDis1D} 
to 2D for \(n_{1,2}\) even yields 
\begin{equation}
\begin{array}{rclcrcl}
\check{f}_{\mu_{1}\mu_{2}}
&=&
\frac{1}{n_{1}n_{2}}
\sum_{i_{1}=1}^{n_{1}}
\sum_{i_{2}=1}^{n_{2}}
\varphi_{\smash{\mu_{1}i_{1}}}^{-}
\varphi_{\!\smash{\mu_{2}i_{2}}}^{-}
f_{i_{1}i_{2}}
\,,
&&
\mu_{1,2}
&=&
1,\ldots,\frac{1}{2}n_{1,2}+1
\,,\\[1mm]
\check{f}_{\mu_{1}i_{2}}
&=&
\check{f}_{\mu_{1}\!1}
+2\mathop{\mathrm{rea}}
\sum_{\mu_{2}=2}^{\frac{1}{2}n_{2}}
\varphi_{\smash{\mu_{2}i_{2}}}^{+}
\check{f}_{\mu_{1}\mu_{2}}
+(-1)^{i_{2}+1}\check{f}_{\mu_{1}\frac{1}{2}n_{2}+1}
\,,
&&
i_{2}
&=&
1,\ldots,n_{2}
\,,\\[1mm]
f_{i_{1}i_{2}}
&=&
\check{f}_{1i_{2}}
+2\mathop{\mathrm{rea}}
\sum_{\mu_{1}=2}^{\frac{1}{2}n_{1}}
\varphi_{\smash{\mu_{1}i_{1}}}^{+}
\check{f}_{\mu_{1}i_{2}}
+(-1)^{i_{1}+1}\check{f}_{\frac{1}{2}n_{1}+1\,i_{2}}
\,,
&&
i_{1}
&=&
1,\ldots,n_{1}
\,,
\end{array}
\label{equ:FieReaSerForDis2D}
\end{equation} 
for \(f\) real-valued via modal periodicity 
and re-indexing \(\mu_{d}=\kappa_{d}+1\), with now 
\begin{equation}
\varphi_{\mu_{d}i_{d}}^{\pm}
:=e^{\pm2\pi\imath(\mu_{d}-1)(i_{d}-1)/n_{d}}
\,.
\end{equation}
Actual calculations employ of course the fast version of 
\eqref{equ:FieReaSerForDis2D}. 

\subsection{The case of plane deformation}

Restricting attention to \((x_{1},x_{2})\)-plane deformation as in the paper, 
\begin{equation}
\mathbf{F}
=\left\lbrack
\begin{array}{ccc}
F_{11}&F_{12}&0
\\
F_{21}&F_{22}&0
\\
0&0&1
\end{array}
\right\rbrack
\,,\quad
\mathbf{E}
=\left\lbrack
\begin{array}{ccc}
E_{11}&E_{12}&0
\\
E_{12}&E_{22}&0
\\
0&0&0
\end{array}
\right\rbrack
\,,\quad
\mathbf{P}
=\left\lbrack
\begin{array}{ccc}
P_{11}&P_{12}&0
\\
P_{21}&P_{22}&0
\\
0&0&P_{33}
\end{array}
\right\rbrack
\,,
\label{equ:DefPlaComCar}
\end{equation}
hold for the deformation gradient, symmetric Green strain, 
and first PK stress, respectively. In particular, the form of 
\(\mathbf{E}\) follows from that of \(\mathbf{F}\) via \Cref{equ:StnGre}, 
and the form of \(\mathbf{P}\) from both of these 
in the context of the isotropic elastic Saint Venant-Kirchhoff 
relation \Cref{equ:ElaKirVenSai}. Likewise, the reduced form 
\begin{equation}
\mathbf{A}=\bar{\mathbf{A}}+\tilde{\mathbf{A}}
\,,\quad
\bar{\mathbf{A}}
=\left\lbrack
\begin{array}{ccc}
\bar{A}_{11}&\bar{A}_{12}&0
\\
\bar{A}_{21}&\bar{A}_{22}&0
\\
0&0&\bar{A}_{33}
\end{array}
\right\rbrack
\,,\ \ 
\tilde{\mathbf{A}}
=\left\lbrack
\begin{array}{ccc}
0&0&\tilde{A}_{13}
\\
0&0&\tilde{A}_{23}
\\
\tilde{A}_{31}&\tilde{A}_{32}&0
\end{array}
\right\rbrack
\,,
\label{equ:DefPlaStsPot}
\end{equation}
of \(\mathbf{A}\) applies in this case, 
consistent with \eqref{equ:DefPlaComCar}${}_{3}$. 
In this context, the fast version of \eqref{equ:FieReaSerForDis2D} 
is applied to each non-zero component of \(\mathbf{P}_{\!i_{1}i_{2}}\) 
and \(\check{\mathbf{P}}_{\!\mu_{1}\mu_{2}}\) in each hidden layer 
of the Pg- and PiFNO architectures in \Cref{fig:ArcANcP}(a). 
In the latter case, \(\check{\mathbf{P}}_{\!\mu_{1}\mu_{2}}^{\mathrm{out}}\) 
determine 
\begin{equation}
\begin{array}{rcl}
\skew4\check{\mathbf{d}}_{\mu_{1}\mu_{2}}^{\mathrm{out}}
&=&
\dfrac{2\pi\imath}{l}
\,\check{\mathbf{P}}_{\!\mu_{1}\mu_{2}}^{\mathrm{out}}
\,(\mu_{1}-1,\mu_{2}-1,0)
\\[3mm]
&=&
\dfrac{2\pi\imath}{l}
\left\lbrack
\begin{array}{c}
(\mu_{1}-1)\check{P}_{11\,\mu_{1}\mu_{2}}^{\mathrm{out}}
+(\mu_{2}-1)\check{P}_{12\,\mu_{1}\mu_{2}}^{\mathrm{out}}
\\
(\mu_{1}-1)\check{P}_{21\,\mu_{1}\mu_{2}}^{\mathrm{out}}
+(\mu_{2}-1)\check{P}_{22\,\mu_{1}\mu_{2}}^{\mathrm{out}}
\\
0
\end{array}
\right\rbrack
\,,\quad
(\mu_{1},\mu_{2})\neq(1,1)
\,,
\end{array}
\end{equation}
in the context of discretized, truncated form of 
\eqref{equ:StsDivRepFouAryCar} for evaluation of 
\(\mathop{\mathrm{div}}\mathbf{P}_{\!a}^{\smash{\mathrm{out}}}\)
in the PiFNO loss function \Cref{equ:FunLosInfPhy}. 
Analogously,  the fast version of 
\eqref{equ:FieReaSerForDis2D} is applied to each non-zero 
component of 
\(\mathbf{A}_{i_{1}i_{2}}\) and \(\check{\mathbf{A}}_{\mu_{1}\mu_{2}}\) 
in each hidden layer of the PeFNO architecture. Lastly, the discretized form of  
\eqref{equ:PotStsRepFouCof}, i.e., 
\begin{equation}
\check{\mathbf{P}}_{\!\mu_{1}\mu_{2}}^{\mathrm{out}}
=\left\lbrace
\begin{array}{lcl}
\check{\mathbf{A}}_{11}^{\mathrm{out}}
&&
(\mu_{1},\mu_{2})=(1,1)
\\[1mm]
\dfrac{2\pi\imath}{l}
\check{\mathbf{A}}_{\mu_{1}\mu_{2}}^{\mathrm{out}}
\lbrack\mathop{\mathrm{axt}}(\mu_{1}-1,\mu_{2}-1,0)\rbrack^{\mathrm{T}}
&&
(\mu_{1},\mu_{2})\neq(1,1)
\end{array}
\right.
\end{equation}
with
\begin{equation}
\begin{array}{l}
\check{\mathbf{A}}_{\mu_{1}\mu_{2}}^{\mathrm{out}}
\lbrack\mathop{\mathrm{axt}}(\mu_{1}-1,\mu_{2}-1,0)\rbrack^{\mathrm{T}}
\\
\quad=\ \ 
\left\lbrack
\begin{array}{ccc}
(\mu_{2}-1)\check{A}_{13\,\mu_{1}\mu_{2}}^{\mathrm{out}}&
-(\mu_{1}-1)\check{A}_{13\,\mu_{1}\mu_{2}}^{\mathrm{out}}&
0
\\
(\mu_{2}-1)\check{A}_{23\,\mu_{1}\mu_{2}}^{\mathrm{out}}&
-(\mu_{1}-1)\check{A}_{23\,\mu_{1}\mu_{2}}^{\mathrm{out}}&
0
\\
0&0&
(\mu_{1}-1)\check{A}_{32\,\mu_{1}\mu_{2}}^{\mathrm{out}}
-(\mu_{2}-1)\check{A}_{31\,\mu_{1}\mu_{2}}^{\mathrm{out}}
\end{array}
\right\rbrack
\,,
\end{array}
\end{equation} 
determines the transformation 
\(\mathbf{A}^{\!\mathrm{out}}\to\mathbf{P}^{\mathrm{out}}\)
in the output layer of \Cref{fig:ArcANcP}(b) for the PeFNO, again 
via component-wise application of \eqref{equ:FieReaSerForDis2D} 
in its fast version. 

\end{document}